\documentclass[fleqn,usenatbib]{mnras}

\usepackage{newtxtext,newtxmath}

\usepackage[T1]{fontenc}

\DeclareRobustCommand{\VAN}[3]{#2}
\let\VANthebibliography\thebibliography
\def\thebibliography{\DeclareRobustCommand{\VAN}[3]{##3}\VANthebibliography}

\usepackage{graphicx}	
\usepackage{amsmath}	
\usepackage{amssymb}	

\usepackage{CJK}
\usepackage[whole]{bxcjkjatype}
\usepackage{CJKutf8}

\newcommand{\kms}{km\,s$^{-1}$}
\newcommand{\nick}{$^{56}$Ni}
\newcommand{\msun}{M$_{\odot}$}

\title[SN\,2021zny]{SN\,2021zny: an early flux excess combined with late-time oxygen emission suggests a double white dwarf merger event}

\author[G. Dimitriadis et al.]{
Georgios Dimitriadis$^{1}$\thanks{E-mail: dimitrig@tcd.ie},
Kate Maguire$^{1}$,
Viraj R. Karambelkar$^{2}$,
Ryan J. Lebron$^{3}$,
Chang Liu\begin{CJK*}{UTF8}{gbsn}
(刘畅)
\end{CJK*}$^{4,5}$,
\newauthor
Alexandra Kozyreva$^{6}$,
Adam A. Miller$^{4,5}$,
Ryan Ridden-Harper$^{7}$,
Joseph P. Anderson$^{8,9}$,
Ting-Wan Chen$^{6,10}$,
\newauthor
Michael Coughlin$^{11}$,
Massimo Della Valle$^{12,13}$,
Andrew Drake$^{2}$,
Llu\'is Galbany$^{14,15}$,
Mariusz Gromadzki$^{16}$,
\newauthor
Steven L. Groom$^{17}$,
Claudia P. Guti\'errez$^{18,19}$,
Nada Ihanec$^{8,16}$,
Cosimo Inserra$^{20}$,
Joel Johansson$^{21}$,
\newauthor
Tom\'as E. M\"uller-Bravo$^{14,15}$,
Matt Nicholl$^{22}$,
Abigail Polin$^{23,24}$,
Ben Rusholme$^{17}$,
Steve Schulze$^{21}$,
\newauthor
Jesper Sollerman$^{25}$,
Shubham Srivastav$^{26}$,
Kirsty Taggart$^{27}$,
Qinan Wang$^{28}$,
Yi Yang\begin{CJK*}{UTF8}{gbsn}
(杨轶)
\end{CJK*}$^{29}$,
David R. Young$^{26}$
\\
$^{1}$School of Physics, Trinity College Dublin, The University of Dublin, Dublin 2, Ireland\\
$^{2}$Cahill Center for Astrophysics, California Institute of Technology, Pasadena, CA 91125, USA\\
$^{3}$Department of Physics and Engineering, University of Scranton, Scranton, PA 18510, USA\\
$^{4}$Department of Physics and Astronomy, Northwestern University, 2145 Sheridan Rd, Evanston, IL 60208, USA\\
$^{5}$Center for Interdisciplinary Exploration and Research in Astrophysics (CIERA), Northwestern University, 1800 Sherman Ave, Evanston, IL 60201, USA\\
$^{6}$Max-Planck-Institut f\"ur Astrophysik, Karl-Schwarzschild-Str. 1, 85748 Garching bei M\"{u}nchen, Germany\\
$^{7}$School of Physical and Chemical Sciences—Te Kura Matu, University of Canterbury, Private Bag 4800, Christchurch 8140, New Zealand\\
$^{8}$European Southern Observatory, Alonso de C\'ordova 3107, Casilla 19, Santiago, Chile\\
$^{9}$Millennium Institute of Astrophysics MAS, Nuncio Monsenor Sotero Sanz 100, Off. 104, Providencia, Santiago, Chile\\
$^{10}$Technische Universit{\"a}t M{\"u}nchen, TUM School of Natural Sciences, Physik-Department, James-Franck-Stra{\ss}e 1, 85748 Garching, Germany\\
$^{11}$School of Physics and Astronomy, University of Minnesota, Minneapolis, MN 55455, USA\\
$^{12}$INAF - Osservatorio Astronomico di Capodimonte, Salita Moiariello 16, 80131 Napoli, Italy\\
$^{13}$Department of Physics, Ariel University, Ariel, Israel\\
$^{14}$Institute of Space Sciences (ICE, CSIC), Campus UAB, Carrer de Can Magrans, s/n, E-08193 Barcelona, Spain\\
$^{15}$Institut d'Estudis Espacials de Catalunya (IEEC), E-08034 Barcelona, Spain\\
$^{16}$Astronomical Observatory, University of Warsaw, Al. Ujazdowskie 4, 00-478 Warszawa, Poland\\
$^{17}$IPAC, California Institute of Technology, 1200 E. California Blvd, Pasadena, CA 91125, USA\\
$^{18}$Finnish Centre for Astronomy with ESO (FINC A), FI-20014 University of Turku, Finland\\
$^{19}$Tuorla Observatory, Department of Physics and Astronomy, FI-20014 University of Turku, Finland\\
$^{20}$Cardiff Hub for Astrophysics Research and Technology, School of Physics \& Astronomy, Cardiff University, Queens Buildings, The Parade, Cardiff, CF24 3AA, UK\\
$^{21}$The Oskar Klein Centre, Department of Physics, Stockholm University, Albanova University Center, SE 106 91 Stockholm, Sweden\\
$^{22}$Birmingham Institute for Gravitational Wave Astronomy and School of Physics and Astronomy, University of Birmingham, Birmingham B15 2TT, UK\\
$^{23}$The Observatories of the Carnegie Institution for Science, 813 Santa Barbara St., Pasadena, CA 91101, USA\\
$^{24}$TAPIR, Walter Burke Institute for Theoretical Physics, 350-17, Caltech, Pasadena, CA 91125, USA\\
$^{25}$Department of Astronomy, The Oskar Klein Center, Stockholm University, AlbaNova, 10691 Stockholm, Sweden\\
$^{26}$Astrophysics Research Centre, School of Mathematics and Physics, Queen's University Belfast, Belfast BT7 1NN, UK\\
$^{27}$Department of Astronomy and Astrophysics, University of California, Santa Cruz, CA 95064, USA\\
$^{28}$Physics and Astronomy Department, Johns Hopkins University, Baltimore, MD 21218, USA\\
$^{29}$Department of Astronomy, University of California, Berkeley, CA 94720-3411, USA
}

\date{Accepted XXX. Received YYY; in original form ZZZ}

\pubyear{2022}

\begin{document}
\label{firstpage}
\pagerange{\pageref{firstpage}--\pageref{lastpage}}
\maketitle

\begin{abstract}

We present a photometric and spectroscopic analysis of the ultra-luminous and slowly evolving 03fg-like Type Ia SN\,2021zny. Our observational campaign starts from $\sim5.3$ hours after explosion (making SN\,2021zny one of the earliest observed members of its class), with dense multi-wavelength coverage from a variety of ground- and space-based telescopes, and is concluded with a nebular spectrum $\sim10$ months after peak brightness. SN\,2021zny displayed several characteristics of its class, such as the peak brightness ($M_{B}=-19.95$ mag), the slow decline ($\Delta m_{15}(B) = 0.62$ mag), the blue early-time colours, the low ejecta velocities and the presence of significant unburned material above the photosphere. However, a flux excess for the first $\sim1.5$ days after explosion is observed in four photometric bands, making SN\,2021zny the third 03fg-like event with this distinct behavior, while its $+313$ d spectrum shows prominent [\ion{O}{i}] lines, a very unusual characteristic of thermonuclear SNe. The early flux excess can be explained as the outcome of the interaction of the ejecta with $\sim0.04\:\mathrm{M_{\odot}}$ of H/He-poor circumstellar material at a distance of $\sim10^{12}$ cm, while the low ionization state of the late-time spectrum reveals low abundances of stable iron-peak elements. All our observations are in accordance with a progenitor system of two carbon/oxygen white dwarfs that undergo a merger event, with the disrupted white dwarf ejecting carbon-rich circumstellar material prior to the primary white dwarf detonation.

\end{abstract}




\section{Introduction} \label{sec:intro}

The remarkable homogeneity in the properties of Type Ia supernovae (SNe Ia) establish them as our best cosmological tools to date. Their peak brightness is strongly correlated with their light curve's shape \citep{Phillips93} and colour \citep{Riess96}, and these strong correlations allow us to standarize them and, by measuring their relative distances, unveil the accelerating expansion of the Universe and the discovery of dark energy \citep{Riess98,Perlmutter99}. SNe Ia, in combination with other local universe distance indicators, such as the period–luminosity relation of Cepheid variable stars \citep{Riess2019ApJ}, the tip of the red giant branch \citep[TRGB;][]{Freedman2019ApJ} and Mira variables \citep{Huang2020ApJ} are also able to constrain the local expansion rate \citep{Riess16,Riess2022ApJ}.

While there is a theoretical and observational consensus that SNe Ia originate from the explosive thermonuclear burning \citep{Hoyle1960ApJ} of a degenerate carbon-oxygen white dwarf (WD) in a binary system \citep{Whelan1973ApJ,Iben1984ApJS,Bloom12}, the nature of its binary companion and the explosion mechanism itself remains unknown, maintaining an active debate on the origins of these events \citep[see reviews of][]{Maoz14,Hoeflich2017hsn,Jha2019NatAs}.

Focusing on the vast majority of SNe Ia, the correlation between their maximum luminosity and their light curve shape \citep[usually parameterized with their magnitude decline for the first 15 days after maximum light, $\Delta m_{15}$;][]{Phillips93} can be explained by the nucleosynthetic yield of $^{56}$Ni, the most abundant radioactive element the exploding WD synthesises \citep[][]{Colgate1969ApJ}, that powers the light curve. For a given ejecta mass \citep[usually the maximum mass a degenerate non-rotating C/O WD can sustain, the Chandrasekhar mass, $\mathrm{M_{Ch}}$,][]{Chandrasekhar31}, smaller/larger amounts of $^{56}$Ni lead to fainter/brighter explosions with shorter/longer timescales. This simple approach has been generally successful in explaining the diversity in the bulk of the SN Ia population, from the subluminous 91bg-like \citep{Filippenko1992AJ} to the bright 91T-like \citep{Phillips1992AJ}, both in their light curve properties \citep{Kasen2007ApJ} and in their spectroscopic ones \citep{Parrent2014}. 

High-cadence and/or untargeted transient surveys performed in recent years, such as the Palomar Transient Factory \citep[PTF;][]{Law2009PASP,Rau2009PASP}, the All-Sky Automated Survey for Supernovae \citep[ASAS-SN;][]{Shappee2014ApJ}, the Distance less than 40 Mpc survey \citep[DLT40;][]{Tartaglia2018ApJ}, the Asteroid Terrestrial-impact
Last Alert System \citep[ATLAS;][]{Tonry2018PASP}, the Panoramic Survey Telescope and Rapid Response System \citep[Pan-STARRS;][]{Chambers2016}, the Young Supernova Experiment \citep[YSE;][]{Jones21ApJ} and the Zwicky Transient Facility \citep[ZTF;][]{Bellm2019PASP,Graham2019PASP,Masci2019PASP,Dekany2020PASP} have started to discover peculiar events.  While these events share many observational characteristics with SNe Ia, they do have distinct photometric (e.g. higher or lower peak luminosities for their decline rate) and/or spectroscopic (e.g. the presence of hydrogen/helium) properties, challenging the canonical paradigm of the thermonuclear scenario \citep[see][for a review]{Taubenberger17}.

One of the most puzzling sub-types of SNe Ia is the so-called 03fg-like SNe Ia, a rare subclass of ultra-luminous and slowly-evolving events. The discovery of the prototype SN\,2003fg \citep{Howell06} revealed a brighter peak luminosity ($M_{B} = -20.09$~mag) for its decline rate ($\Delta m_{15}(B) = 0.82$~mag), and using simple analytical models \citep{Arnett82,Jeffery1999} an estimate of the nickel and ejecta mass of more than the Chandrasekhar mass was obtained. Over the next years, and as more 03fg-like SNe Ia were discovered, an intrinsic diversity in the sub-population has been unveiled, with some of them being less luminous and/or faster evolving, or showing a rapid fading in the optical bands with simultaneous increase of the near-infrared (NIR) flux. Moreover, varying spectroscopic properties were found, such as the strengths and velocities of silicon (an element probing the synthesized material in the ejecta) and carbon (an element probing the unburned pristine material from the WD), or the potential presence of oxygen in late-time spectra (see \citealt{Hicken07ApJ,Maeda09ApJ,Scalzo10,Taubenberger11,Chakradhari14MNRAS,Parrent16MNRAS,Taubenberger19,Chen2019ApJ,Hsiao20ApJ,Lu2021ApJ,Dimitriadis2022ApJ} for studies on individual events and \citealt{Taubenberger2013MNRAS,Ashall2021ApJ} for sample studies). 

Initial suggestions for solving the mass puzzle of 03fg-like SNe Ia were rapidly spinning WDs as the progenitors, as the differential rotation can form systems with super-$\mathrm{M_{Ch}}$ masses \citep[][]{Yoon05AA}, leading to the adoption of the `super-Chandrasekhar-mass' moniker. However, these approaches were disputed by numerical simulations \citep{Pfannes10AA,Pfannes10AA2,Hachinger2012MNRAS,Fink18AA}, particularly the nucleosynthesis and the energetics, as they produce substantial amounts of burned material at high ejecta velocities, in contrast with observations of (most of) 03fg-like SNe Ia. Moreover, a super-$\mathrm{M_{Ch}}$ $^{56}$Ni explosion (needed to reproduce the enormous peak luminosity) with low ejecta velocities, will require a huge amount of ejecta mass, leading to strong $\gamma$-ray trapping and bright late-time bolometric light curves, in contrast with the observations \citep{Taubenberger2013MNRAS}. Evidently, the observed properties of 03fg-like SNe Ia, from the early rise to the nickel decay tail, cannot consistently be explained by any $^{56}$Ni – ejecta mass combination, which led to the introduction of alternative scenarios, where the luminosity of the SN is not solely powered by the $^{56}$Ni decay. This can be achieved by the interaction of the ejecta with circumstellar material (CSM) in the close vicinity of the explosion site, that would naturally increase the luminosity at early times and decelerate the ejecta, sustaining a broad light curve \citep[][]{Hicken07ApJ,Scalzo10,Taubenberger11}. The origin of this H-free CSM (as no hydrogen has ever been observed in any 03fg-like SN Ia) is still under debate, with the disrupted secondary C/O WD in a binary WD merger \citep{Raskin13,Raskin14} or the carbon-rich envelope of an asymptotic giant branch (AGB) star at the end of its evolution under the `core-degenerate' scenario \citep{Hoeflich96ApJ,Kashi11MNRAS,Hsiao20ApJ,Ashall2021ApJ} being the primary candidates. However, the main problem with these scenarios is that no clear signatures of this interaction have been observed, either in spectra, as narrow emission lines, or in the light curve evolution, as a deviation of the smooth early-time rise predicted for an explosion in a CSM-free environment.

While the aforementioned observables have never been seen in 03fg-like SNe Ia, various other subtypes of SNe Ia display properties that indicate a different underlying explosion mechanism and/or progenitor system compared to normal SNe Ia. From the spectral side, contrary to normal SNe Ia \citep[e.g.][]{Tucker2020MNRAS}, the peculiar-Ia class of SNe Ia-CSM \citep[][]{Silverman13ApJS} shows narrow H$\alpha$ lines, including at early times, consistent with the presence of a non-degenerate companion's dense H-rich CSM, and occupy a similar area in the absolute magnitude -- $\Delta m_{15}$ parameter space as 03fg-like SNe Ia. On the other hand, normal and underluminous events such as SNe\,2015cp \citep{Graham2019ApJ}, 2016jae \citep{EliasRosa2021AA}, 2018cqj \citep{Prieto2020ApJ} and ASASSN-18tb \citep{Kollmeier2019MNRAS} only revealed narrow $H\alpha$ at later times, supporting a delayed ejecta-CSM interaction scenario.

The situation appears more complicated within the early photometric evolution. Individual nearby normal SNe Ia, observed moments after explosion, such as SNe\,2011fe \citep{Nugent11,Bloom12}, 2014J \citep{Goobar14ApJ} and ASASSN-14lp \citep{Shappee2016ApJ} show a smooth early rise, usually parametrized as a power law, $L\propto t^{\alpha}$, where $\alpha=2$ corresponds to the canonical `expanding fireball' model \citep{Arnett82,Riess99AJ}. Continuous, high cadence observations with transiting exoplanet surveys, such as {\it Kepler/K2} \citep{Olling15Natur,Wang2021ApJ} and {\it Transiting Exoplanet Survey Satellite} \citep{Fausnaugh2021ApJ} find similar results. Statistical sample studies \citep{Conley2006AJ,Hayden10ApJ,Ganeshalingam11MNRAS,GonzalezGaitan12,Firth2015MNRAS,Papadogiannakis2019MNRAS,Miller2020ApJ} find mean values of $1.8\le\alpha\le2.4$, however, many individual SNe in the samples are incompatible with $\alpha=2$.

Next to these well-behaved SNe Ia, some individual objects are clearly inconsistent with a smooth rising light curve, showing early flux excesses of various strengths, timescales and colours. Most notably, a blue and relatively long ($\sim$2 to 5 d) `bump' in the early light curves of SN\,2012cg \citep{Marion16ApJ}, iPTF14atg \citep{Cao2015Natur}, SN\,2017cbv \citep{Hosseinzadeh17ApJ}, SN\,2018oh \citep{Dimitriadis2019ApJ,Shappee2019ApJ,Li2019ApJ} and SN\,2021aefx \citep{Hosseinzadeh2022ApJ}, has been attributed to ejecta interaction with (main sequence or subgiant) non-degenerate companions. However, none of the above SNe Ia have shown signs of stripped material from the donor in late-time spectral observations \citep[e.g. see][]{Maguire2016MNRAS,Shappee2018ApJ,Dimitriadis2019ApJ2,Tucker2019ApJ,Sand2021ApJ}, leading to alternative explanations of the early `bumps', such as the presence of $^{56}$Ni near the surface due to mixing \citep{Piro13ApJ,Magee2020AA} or the production of radioactive material in the ashes of the helium shell under a double-detonation explosion of a sub-$\mathrm{M_{ch}}$ WD \citep{Polin2019ApJ}. The presence of excess nucleosynthetic material in the outermost layers of the ejecta has been proposed for the short-term ($\sim$0.5 days) and redward evolution of SN\,2018aoz's early flux excess \citep{Ni2022NatAs}, with \citet{Jiang2017Natur} and \citet{De2019ApJ} favouring a double-detonation for the longer-lasting red `bumps' of MUSSES1604D and SN\,2018byg, respectively. Finally, \citet{Miller2020ApJ2} and \citet{Burke2021ApJ} identify a long ($\sim$3.5 days) and ultraviolet (UV) bright flux excess for SN\,2019yvq, for which \citet{Siebert2020ApJ}, based on the strong calcium emission at the nebular spectrum, favour a double-detonation origin, although, as \citet{Tucker2021ApJ} note, there is no single explosion model that can simultaneously explain its early- and late-time properties, a situation that is encountered in almost all SNe Ia with early flux excesses \citep{Magee2020AA}. Nevertheless, sample studies of SNe Ia, dedicated to identify these early `bumps', show an intrinsic rate of $18\pm11$ per cent of early flux excesses in SNe Ia \citep{Deckers2022MNRAS,Burke2022arXiv220707681B,Burke2022arXiv220811201B}, posing additional challenges on their interpretation.

Recently, a short-lived flash of optical emission was observed for two overluminous SNe Ia, 2020hvf \citep{Jiang2021ApJ} and 2022ilv \citep{Srivastav2023ApJ}. For SN\,2020hvf, the flux excess was observed during the high cadence Tomo-e Gozen transient survey, using the camera's clear filter, and lasted for $\sim1$ day, while for SN\,2022ilv, observations in the ATLAS $\textit{o}$-band showed a similar early time behaviour. The authors modeled the rising light curves, and favor interaction of the ejecta with a CSM mass of $\sim10^{-2}-10^{-3}\:\mathrm{M_{\odot}}$ at an outer edge radius of $\sim10^{13}$ cm. While SN\,2020hvf has some notable spectral differences compared to 03fg-like SNe Ia (relatively weak carbon lines and extremely high ejecta velocities), these two events provide the first detection of a flux excess for members of the 03fg-like subclass.

In this paper, we present observations of SN\,2021zny, an 03fg-like SN Ia, discovered $\sim$hours after explosion, classified two weeks before maximum brightness and densely monitored with ground- and space-based facilities. Our $\sim$10 months of multi-wavelength photometric and spectroscopic coverage makes SN\,2021zny one of the most well-observed  03fg-like SNe Ia, for which we identify two striking features. Firstly, an early, short-lived flash is observed in four photometric filters, which is consistent with a small amount of H-free CSM interacting with the SN ejecta and secondly, the detection of oxygen in its $+313$d late-time spectrum. We present the discovery of SN\,2021zny, our observational campaign and the techniques we used for the reduction of our data in Section~\ref{sec:obs_data_red}. The analysis of its photometric and spectroscopic properties, alongside a discussion on its distance and extinction along the line of sight is presented in Section~\ref{sec:analysis}. We discuss our findings in the context of the proposed progenitor systems of 03fg-like SNe Ia in Section~\ref{sec:discussion}, and, finally, conclude in Section~\ref{sec:conclusion}.

Throughout this paper, we will use the moniker 03fg-like SNe Ia to describe the members of this peculiar SN Ia subclass, noting that various monikers have been used in the literature, such as `super-Chandrasekhar-mass' SNe Ia (SC SNe Ia), 09dc-like and (carbon-rich) overluminous SNe Ia. Moreover, every phase of a light curve is in rest-frame days. Finally, we adopt the AB magnitude system and a Hubble constant of $H_0 = 73$ km s$^{-1}$ Mpc$^{-1}$.

\section{Discovery, Observations and Data Reduction} \label{sec:obs_data_red}

In this Section, we present the discovery of SN\,2021zny, its classification and our photometric and spectroscopic followup observations.

\subsection{Discovery and Classification} \label{sec:discovery}

SN\,2021zny was discovered on UT 2021 September 22.37 by the Zwicky Transient Facility \citep[ZTF;][]{Bellm2019PASP,Graham2019PASP,Masci2019PASP,Dekany2020PASP}, with the internal survey name ZTF21acdmwae, and reported on the Transient Name Server (TNS\footnote{\url{https://www.wis-tns.org/}}) on UT 2021 September 26.51 \citep{ZTF2021TNSTR3311}, with a discovery magnitude of $\textit{r}=19.33$~mag. Forced photometry on images taken by ZTF prior to discovery revealed that the SN was also present in previous epochs, with our first detection being on UT 2021 September 19.50 ($\mathrm{MJD}=59476.50$) at $\textit{g}=20.35\pm0.17$~mag (with non-detections in $\textit{r}$ and $\textit{i}$ down to $20.94$ and $20.50$~mags, respectively). Our last non-detection in both $\textit{g}$ and $\textit{r}$-bands was on UT 2021 September 17.4 at $>21.85$ and $21.42$~mags, respectively.

\begin{figure}
\includegraphics[width=\columnwidth]{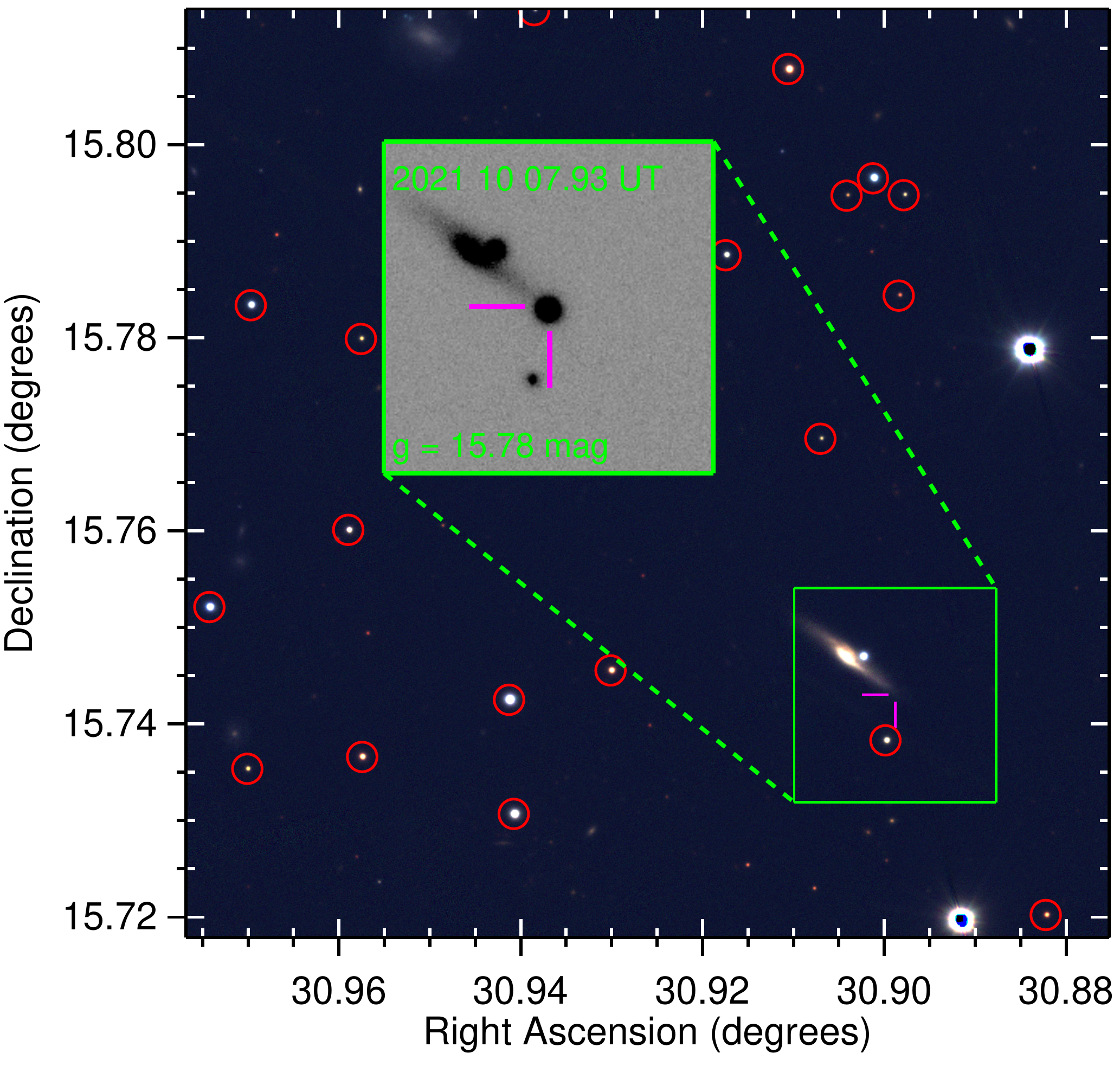}
\caption{Pan-STARRS 5\arcmin{}$\times$5\arcmin{} colour composite (\textit{g}/\textit{r}/\textit{i}) image stamp of the field of CGCG 438-018, the host of SN\,2021zny. The location of the SN is indicated with the magenta tick-marks. Standard stars in the field, used for calibration of our photometry, are marked with red circles. The green inset shows a zoomed-in (80\arcsec{}$\times$80\arcsec{}) region, centered on the SN location , taken at a phase of $-3.4$ d from $\textit{B}$-band maximum with LCO $\textit{g}$-band.}
\label{fig:image}
\end{figure}

The host of SN\,2021zny is CGCG 438-018, an edge-on (star-forming) galaxy, with the SN located at $\alpha=02^{\rm{h}}03^{\rm{m}}35^{\rm{s}}.800$, $\delta=+15^{\rm{o}}44\arcmin33\arcsec.36$ (J2000.0), $18.90$\arcsec\ West and $14.93$\arcsec\ South of its host galaxy's core, along its dust lane. We present a deep pre-explosion Pan-STARRS colour composite ($\textit{g}$/$\textit{r}$/$\textit{i}$) image stamp of CGCG 438-018 with the location of SN\,2021zny marked with magenta tick-marks in Fig.~\ref{fig:image}, and the green inset showing a zoomed-in region of an LCO $\textit{g}$-band image of the supernova, taken at $-3.4$ days from \textit{B}-band maximum.

SN\,2021zny was classified as a young ($\sim$8 days before \textit{B}-band maximum) 03fg-like SN Ia based on an optical spectrum obtained on UT 2021 September 29 by \citet{Yamanaka_class} with the KOOLS-IFU attached to the 3.8-m Seimei telescope at the Okayama Observatory. An additional spectrum obtained from ZTF two days before with the Double Spectrograph (DBSP) mounted on the 5.1-m (P200) Hale Telescope at the Palomar Observatory \citep{Oke1982PASP} confirmed the classification, as both spectra showed a deep absorption feature at $\sim$6,300~\AA. Such a feature can be attributed to \ion{C}{ii} $\lambda$6580 at a similar velocity to one of the most characteristic broad, \ion{Si}{ii}$\lambda$6355 absorption line centered near $\sim$6,150~\AA. This classification, alongside our extremely early detection of SN\,2021zny led us to initiate an extensive follow-up campaign.

\subsection{Observations and Data Reduction} \label{sec:data}

The majority of our photometric and spectroscopic data were obtained within the ZTF collaboration, with additional observations from various other telescopes and instruments. In the next sections, we present the data and the reduction techniques performed.

\subsubsection{Photometry} \label{sec:phot}

SN\,2021zny was observed with ZTF's wide-field camera (in \textit{g,r} and \textit{i}-band filters) mounted on the $1.2$-m Samuel Oschin (P48) Telescope, with dense coverage from $-21.4$ to $+40.8$ d and from $+120.4$ to $+127.3$ d from peak brightness. The images were processed with the  pipeline as described in \citet{Masci2019PASP}, which produces difference imaging, forced-photometry calibrated light curves and post-processed, for quality filtering, with the methods described in \citet{Yao2019ApJ}.

Additional optical photometry was obtained with the IO:O camera of the Liverpool Telescope \citep[LT;][]{Steele2004SPIE} in \textit{u,g,r,i} and \textit{z}-band filters (PL21A09, PI: Deckers), with the Asteroid Terrestrial-impact Last Alert System \citep[ATLAS;][]{Tonry2018PASP} in the orange band, and with the Sinistro cameras of the Las Cumbres Observatory \citep[LCO;][]{LCOGT13PASP} network of 1-m telescopes through ePESSTO+ \citep{Smartt2015AA} OPTICON time (2021B/001, PI: Inserra) in \textit{u,g,r} and \textit{i}-band filters. The LT images were reduced with the IO:O pipeline\footnote{\url{https://telescope.livjm.ac.uk/TelInst/Inst/IOO/}}. LT images were subtracted against Pan-STARRS \citep[PS;][]{Tonry2012} reference imaging, and PSF photometry was performed and calibrated against PS photometric standards. The ATLAS photometry was obtained from the ATLAS forced photometry server \citep{Shingles2021TNSAN}. The Sinistro images were processed with a dedicated \textsc{python/pyraf} pipeline\footnote{\url{https://github.com/LCOGT/lcogtsnpipe}}. UV photometric observations were performed with the Ultraviolet Optical Telescope \citep[UVOT;][]{Roming05} on board the {\it Neil Gehrels Swift Observatory} \citep{Gehrels04}, with reference images taken on UT 2022 November 26. NIR photometric (\textit{J-,H-} and \textit{K-} band) observations were obtained with  SofI \citep{Moorwood1998} on the 3.58-m New Technology Telescope (NTT) through ePESSTO+ \citep{Smartt2015AA} and the Wide Field Infrared Camera \citep[WIRC;][]{Wilson2003SPIE} on P200. The WIRC images were reduced using a custom \textsc{PYTHON} package, that uses the \texttt{Swarp} \citep{Bertin2002} and \texttt{Scamp} \citep{Bertin2006} packages. 

\begin{figure}
\includegraphics[width=\columnwidth]{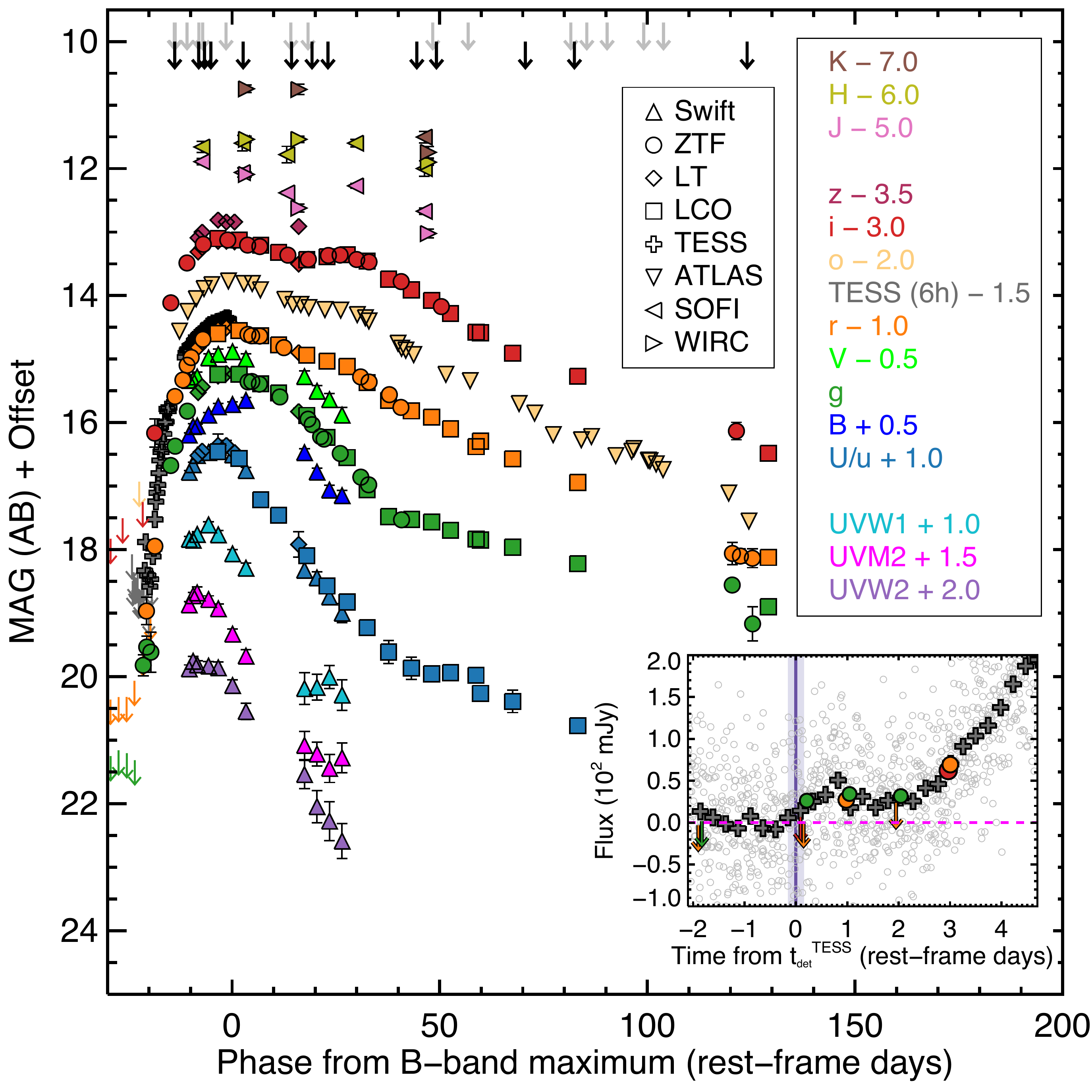}
\caption{Multi-wavelength (\textit{UV}+\textit{uBgVroiz}+\textit{JHK}+\textit{TESS}) light curves of SN\,2021zny in rest-frame days, with respect to \textit{B}-band maximum. Downward arrows mark non-detections at the location of the SN. The light curves are plotted with different symbols/colours and offset, as described in the legends. Downward arrows at the top of the figure correspond to the phases of our $-14$ d to $124$ d spectroscopic series (Fig.~\ref{fig:spec_series} and Table~\ref{tab:spec_log}), with grey (black) colours referring to low (medium) resolution spectra, respectively. The \textit{gri}+\textit{TESS} light curves during the first 4 days after $t_{\mathrm{det}}^{TESS}$ (see Section~\ref{sec:rising_lc}) are shown in the inset, where open grey circles correspond to the raw \textit{TESS} data.}
\label{fig:phot}
\end{figure}

Finally, the field of CGCG 438-018 was serendipitously observed by the {\it Transiting Exoplanet Survey Satellite} \citep[{\it TESS};][]{Ricker2015JATIS}, during Year 4 of the survey, at Sector 43, from UT 2021 September 16 until October 12, covering the rise up to $\sim$1 day before peak brightness. {\it TESS} data were reduced and calibrated with \texttt{TESSreduce}\footnote{\url{https://github.com/CheerfulUser/TESSreduce}}, a dedicated pipeline optimised for SN photometry \citep{Ridden-Harper2021arXiv}. For additional information on the  reduction steps we refer to \citet{Tinyanont2022MNRAS}. Alongside reduction we also verified the validity of the \textit{TESS} signal as \textit{TESS} data are subject to strong systematics around times of intense scattered background light. In this case SN\,2021zny fell on a region of \textit{TESS} detector known as a ``strap" which effectively enhances the quantum efficiency of strap pixels, and complicates the background subtraction. While the flux excess occurs before the scattered light, which begins at $\sim$59483~MJD, we check test apertures within the strap pixels at varying distances from the SN\,2021zny aperture and find no features concurrent with the excess flux. We also verify that the signal is present in all pixels used in the \textit{TESS} aperture, and therefore is unlikely to be a systematic of a single pixel. Our complete photometric data set is presented in Tables~\ref{tab:phot_table}, \ref{tab:phot_table_swift} and \ref{tab:phot_table_tess}, and shown in Fig.~\ref{fig:phot}.

\subsubsection{Spectroscopy} \label{sec:spec}

Spectroscopic observations of SN\,2021zny were initiated immediately after discovery, resulting in a dense and wide coverage, spanning from $-13.9$ to $+117.5$ d from \textit{B}-band maximum. We obtained a total of 15 low resolution spectra: 11 with SED Machine \citep[SEDM;][]{Blagorodnova2018PASP}, a fully-filled integral field spectrograph mounted on the 1.5-m Palomar 60-inch Telescope (P60), and 4 with the SPectrograph for the Rapid Acquisition of Transients \citep[SPRAT;][]{Piascik2014SPIE} on the Liverpool Telescope (PL21A09, PI: Deckers). We additionally obtained 17 medium resolution spectra: two with DBSP, seven with the ESO Faint Object Spectrograph and Camera \citep[EFOSC2;][]{Buzzoni1984Msngr} on NTT through ePESSTO+ \citep{Smartt2015AA}, one with the Low-Resolution Imaging Spectrometer \citep[LRIS;][]{LRIS} on the Keck I telescope, one with the Low Dispersion Survey Spectrograph 3 (LDSS3)\footnote{\url{https://www.lco.cl/technical-documentation/index-2/}} on the Magellan II telescope and two with the Alhambra Faint Object Spectrograph and Camera (ALFOSC)\footnote{\url{http://www.not.iac.es/instruments/alfosc}} at the 2.56-m Nordic Optical Telescope (NOT) at the Observatorio del Roque de los Muchachos on La Palma (Spain). A late-time spectrum ($313$ d from \textit{B}-band maximum) was obtained with the DEep Imaging Multi-Object Spectrograph \citep[DEIMOS;][]{Faber2003SPIE} on the Keck II telescope.

The SEDM spectra were reduced with \textsc{pysedm}\footnote{\url{https://github.com/MickaelRigault/pysedm}} \citep{Rigault2019AA,Kim2022PASP}. All long-slit spectral observations were reduced using standard \textsc{iraf/pyraf}\footnote{IRAF is distributed by the National Optical Astronomy Observatory, which is operated by the Association of Universities for Research in Astronomy (AURA) under a cooperative agreement with the National Science Foundation.} and IDL/python routines for bias subtractions and flat fielding of the two-dimensional spectral images. The wavelength solution was derived using arc lamps and additionally verified against bright night-sky emission lines, while the final flux calibration and (whenever possible) telluric lines removal were performed using spectro-photometric standard star spectra, obtained on the same night. We used dedicated pipelines for the EFOSC2\footnote{\url{https://github.com/svalenti/pessto}}, the ALFOSC\footnote{\url{https://github.com/jkrogager/PyNOT}} and the DEIMOS\footnote{\url{https://pypeit.readthedocs.io/en/latest/}} data. Finally, spectrophotometry of the final flux-calibrated spectra was compared with broad-band photometry of the same night (or interpolated values if no imaging was performed) and scaled with a constant value if necessary. 

\begin{figure}
\includegraphics[width=\columnwidth]{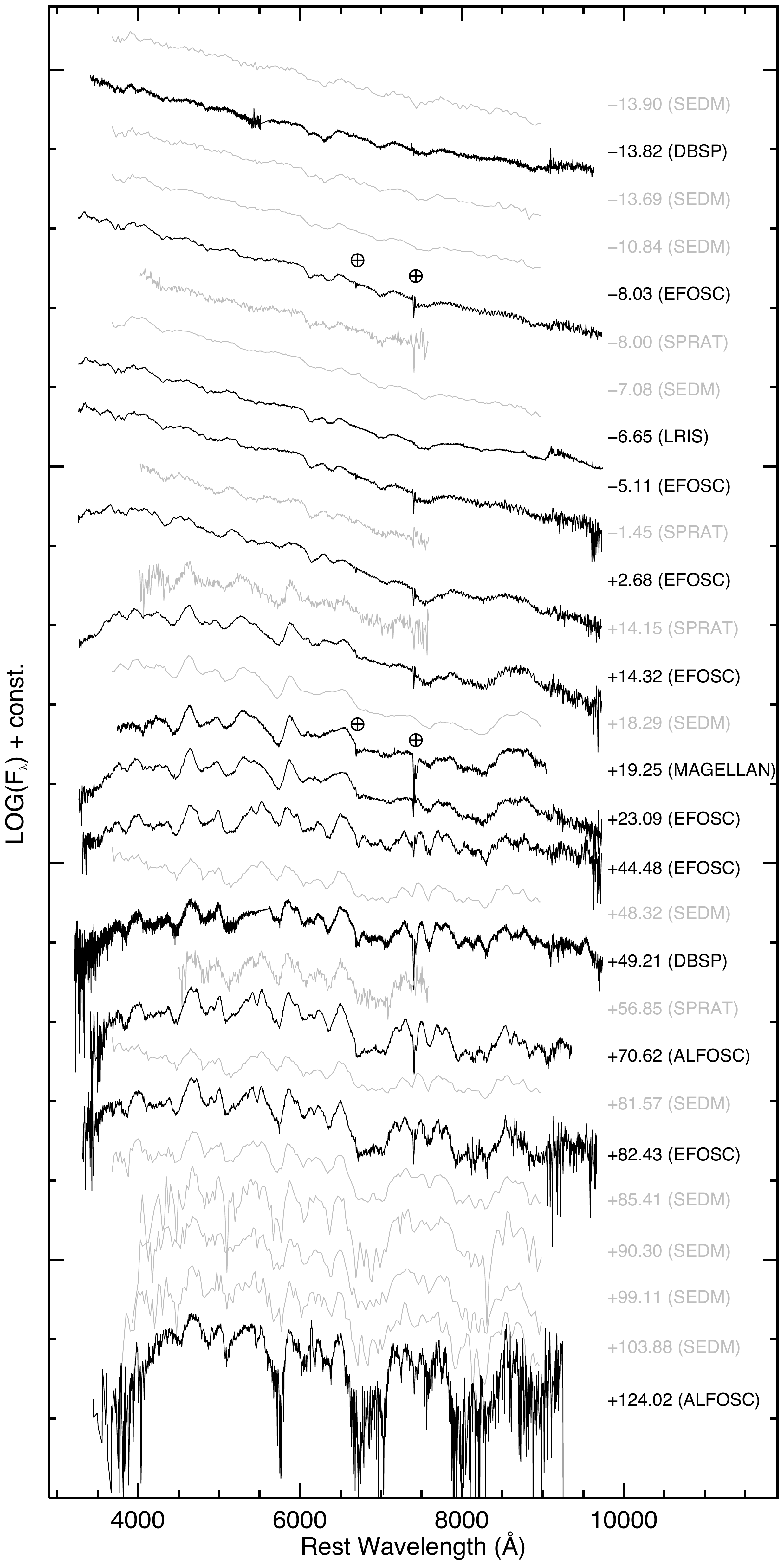}
\caption{The $-14$ to $+124$ d spectroscopic series of SN\,2021zny. The spectra are presented in black (grey) colours, corresponding to medium (low) resolution spectroscopic observations. Its spectrum phase and source is additionally labelled. The spectra have been corrected for Milky Way and host galaxy extinction (see Section~\ref{sec:dist_extinction}) and placed in rest-frame wavelength. Telluric features are marked with the Earth symbol. Detailed information of each observation is provided in Table~\ref{tab:spec_log}.}
\label{fig:spec_series}
\end{figure}

Fig.~\ref{fig:spec_series} shows our $-14$ to $+124$ d spectral series, with detailed information of each observation provided in Table~\ref{tab:spec_log}. The $+313$ d spectrum will be presented and analysed in Section~\ref{sec:spec_analysis}. The complete spectroscopic data set is available in the electronic edition.

\section{Analysis} \label{sec:analysis}

In this Section, we discuss SN\,2021zny's distance and extinction along the line of sight, present its maximum light photometric and spectroscopic properties and focus on the early rise of its light curve.

\subsection{Distance and Extinction} \label{sec:dist_extinction}

NED\footnote{\url{https://ned.ipac.caltech.edu/}} reports four redshift-independent distances for CGCG 438-018, with a mean value of their distance moduli of $35.19\pm0.16$ mag. The reported redshift of the galaxy is $z = 0.026602 \pm 0.000064$ \citep{2005ApJS}, and the cosmological distance of the host, assuming H$_{0} = 73.0 \pm 5.0$~km~s$^{-1}$~Mpc$^{-1}$ and correcting for peculiar motions related to the Virgo cluster and Great Attractor \citep{Mould00}, is estimated at $D = 106.6 \pm 7.5$~Mpc (distance modulus of $DM = 35.14 \pm 0.15$ mag). At this distance, SN\,2021zny is located $12.45$ kpc from the galaxy's core. We will use the cosmological distance in our analysis in order to facilitate better comparisons with other SNe.

Using the \citet{Schlafly11} dust maps, we recover $E(B-V)_{\rm MW}=0.0445\pm0.0015$~mag, which corresponds to a Milky Way extinction along the line of sight $A(V)_{\rm MW}\simeq0.14$~mag of visual extinction. For the host galaxy extinction, we use our LRIS spectra to estimate the equivalent width of the \ion{Na}{i} D absorption line at the host's redshift and find a value of $0.74\pm0.25$~\AA. Using equation 9 from \citet{Poznanski2012MNRAS}, we infer $E(B-V)_{\rm host}=0.10\pm0.07$~mag, which corresponds, assuming a \citet{Fitzpatrick99} reddening law with $R_{V}=3.1$, to a visual extinction $A(V)_{\rm host}\sim0.3$~mag. Thus, the total amount of reddening on the line of sight is $E(B-V)_{\rm total}=0.14\pm0.07$~mag and we adopt this value to correct all of our photometry and spectra.

\subsection{Photometric Properties} \label{sec:phot_analysis}

SN\,2021zny peaked in \textit{B}-band on 2021 October 11.77 UT ($\mathrm{MJD_{\textit{B},peak}}=59498.46\pm0.5$), determined by low-order polynomial fits to our {\it Swift} \textit{B}-band light curve. The observed peak \textit{B}-band magnitude was $B = 15.79\pm0.04$~mag, and taking into account the distance to CGCG 438-018 and the extinction on the line of sight (Section~\ref{sec:dist_extinction}), we estimate the \textit{B}-band peak absolute magnitude to be $-19.95\pm0.17$~mag. The magnitude decline in \textit{B}-band after 15 d was $\Delta m_{15}(B) = 0.62\pm0.09$~mag.

\begin{figure}
\begin{center}
\includegraphics[width=\columnwidth]{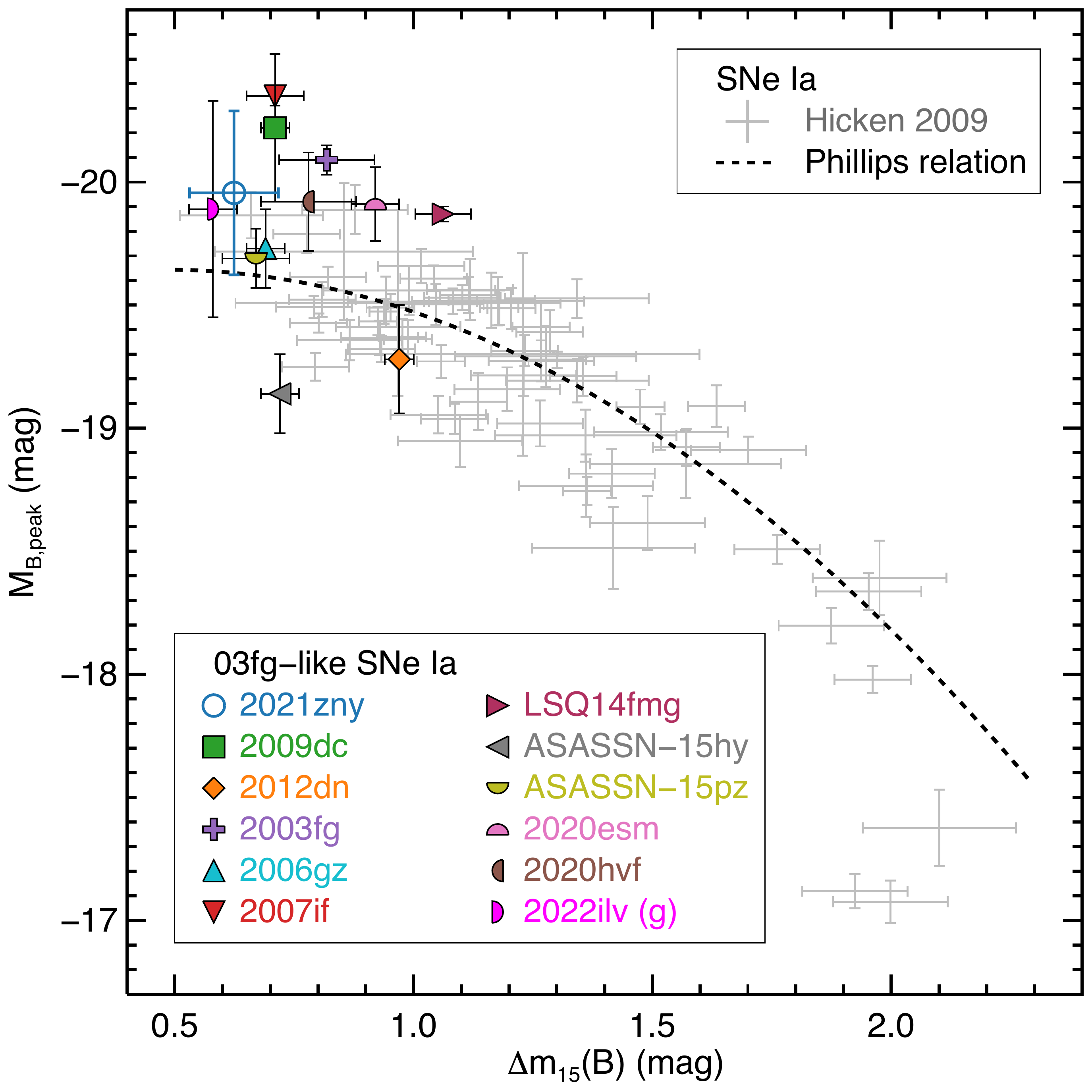}
\caption{The absolute $B$-band magnitude vs. $\Delta m_{15}(B)$ of the CfA SN Ia sample from \citealt{Hicken09ApJ} (grey crosses). The dashed line corresponds to the Phillips relation. We overplot various well-studied 03fg-like SNe Ia as described in the legend. Note that for SN 2022ilv we plot $g$-band measurements.}
\label{fig:phillips}
\end{center}
\end{figure}

In Fig.~\ref{fig:phillips}, we show the absolute $B$-band peak magnitude against $\Delta m_{15}(B)$ for a sample of well-observed 03fg-like SNe Ia, alongside the CfA SN Ia sample from \citet{Hicken09ApJ}. The intrinsic diversity of the 03fg-like SN Ia population is evident in this parameter space, with next to the usual population of high luminosity and slowly evolving SNe 2003fg \citep{Howell06}, 2007if \citep{Scalzo10}, 2009dc \citep{Taubenberger11}, 2020hvf \citep{Jiang2021ApJ} and 2022ilv \citep{Srivastav2023ApJ} lying the slightly dimmer 2006gz \citep{Hicken07ApJ} and ASASSN-15pz \citep{Chen2019ApJ}, the slightly dimmer but faster evolving LSQ14fmg \citep{Hsiao20ApJ} and 2020esm \citep{Dimitriadis2022ApJ}, the faint and faster evolving 2012dn \citep[][]{Taubenberger19} and the faint and slow evolving ASASSN-15hy \citep{Lu2021ApJ}. SN\,2021zny lies on the extreme end of the magnitude-decline range, with $\Delta m_{15}(B)$ values similar to the slow evolving SNe 2006gz, 2007if and 2009dc, with its peak luminosity being $\sim0.33$ mag brighter than what is expected for its decline rate.

\begin{figure*}
\begin{center}
\includegraphics[width=0.97\textwidth]{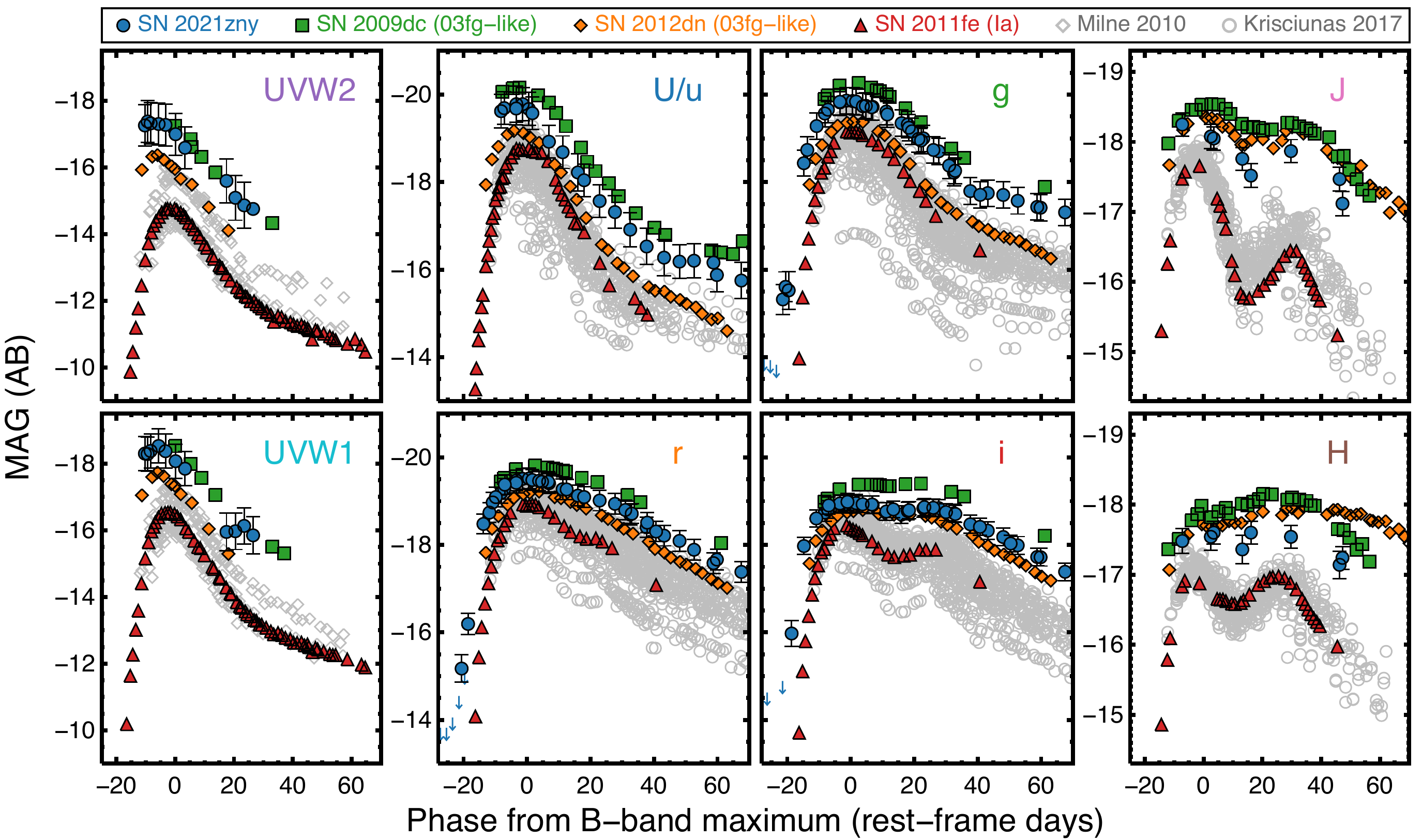}
\caption{The UV ({\it Swift} \textit{UVW2} and \textit{UVW1}-band), optical (\textit{u,g,r,i}-band, various instruments) and NIR (\textit{J,H}-band, various instruments) absolute magnitude light curves of SN\,2021zny (blue circles). Comparison samples in the UV \citep[open grey diamonds;][]{Milne2010ApJ} and the optical/NIR \citep[open grey circles;][]{Krisciunas2017AJ} are additionally shown, alongside the two well observed 03fg-like SNe 2009dc \citep[green squares; data from][]{Taubenberger11,Brown2014,Friedman2015ApJS} and 2012dn \citep[orange diamonds; data from][]{Taubenberger19,Brown2014,Yamanaka2016PASJ}, and the normal SN Ia 2011fe \citep[red upward triangles; data from][]{Pereira13,Brown2012ApJ,Matheson2012ApJ}.}
\label{fig:abs_phot}
\end{center}
\end{figure*}

Fig.~\ref{fig:abs_phot} shows the light curves of SN\,2021zny in absolute magnitudes. We compare the UV, optical and NIR light curves with literature samples of low-redshift, normal SNe Ia from \citet{Milne2010ApJ} and \citet{Krisciunas2017AJ} in similar filters, and with two well-observed 03fg-like SNe Ia 2009dc and 2012dn (which additionally represent the two extremes in the 03fg-like population), and the well-observed normal SN Ia 2011fe. While all of these comparison SNe were photometrically observed in the \textit{UBVRI}-bands, they all have excellent spectrophotometric coverage, making it possible to estimate \textit{ugri}-band light curves, suitable for direct comparison. SN\,2021zny's \textit{gri}-band light curves have additionally been k-corrected based on SN 2009dc. Our SN\,2021zny spectral series (Fig.~\ref{fig:spec_series} and Table~\ref{tab:spec_log}) consists of mainly low-resolution spectra, but it is dense enough in time to compare the estimated k-corrections per filter and as a function of time with the k-corrections of SNe 2009dc \citep{Taubenberger11}, 2012dn \citep{Taubenberger19} and (as a cross-check) 2011fe \citep{Pereira13}. The similarities of the k-corrections with the SN 2009dc ones, alongside the fact that the two SNe are close in the absolute magnitude -- decline rate parameter space, led us to use the 2009dc k-corrections at the corresponding epochs (from $-9$ to $100$ d from maximum light). We do not attempt to extrapolate at phases $<-9$ d, a crucial period, due to the presence of the flux excess, however, we note that for the earliest SN 2009dc spectrum ($-9$ days from peak), the k-corrections are $\sim -0.05, -0.11$ and $-0.09$ for \textit{gri}, respectively, which are considerably smaller than the observed photometric uncertainties.

The light curve evolution of SN\,2021zny demonstrates the usual photometric properties associated with 03fg-like SNe Ia. Compared to the normal SN 2011fe, SN\,2021zny is substantially brighter at peak ($\sim$ 2.5, 1 and 0.7 mag in the UV, optical and NIR, respectively), with a slower evolution in all photometric bands. This behaviour is especially evident in the UV: our earliest {\it Swift} observations were taken $\sim$ 10 d before \textit{B}-band maximum (when SN\,2021zny was already 3.9 and 3.1 mag brighter than SN 2011fe in \textit{UVW2}- and \textit{UVW1}-band, respectively), and SN\,2021zny shows a nearly flat evolution, indicating that it peaked considerably earlier and brighter than SN 2011fe. Moreover, SN\,2021zny lacks any clear evidence of the (distinctive in normal SNe Ia) strong secondary maximum in the \textit{r,i} and NIR photometric bands. Overall, the photometric evolution of SN\,2021zny strongly resembles SN 2009dc although somewhat fainter, suggesting similar conditions in the SN ejecta with a lower synthesized $^{56}$Ni mass.

The most striking characteristic of SN\,2021zny is its light curve evolution from $\sim-$21 to $-$18 d from maximum light (see inset of Fig.~\ref{fig:phot}). All of our photometric observations at those epochs (in the \textit{gri}- and \textit{TESS}-bands) show a prominent flux excess relative to what is expected for the majority of normal SN Ia explosions. However, such an early flux excess exhibits a notable resemblance to that of SN Ia 2020hvf. We will discuss in more detail this remarkable behaviour in Section~\ref{sec:rising_lc}.

\begin{figure}
\begin{center}
\includegraphics[width=\columnwidth]{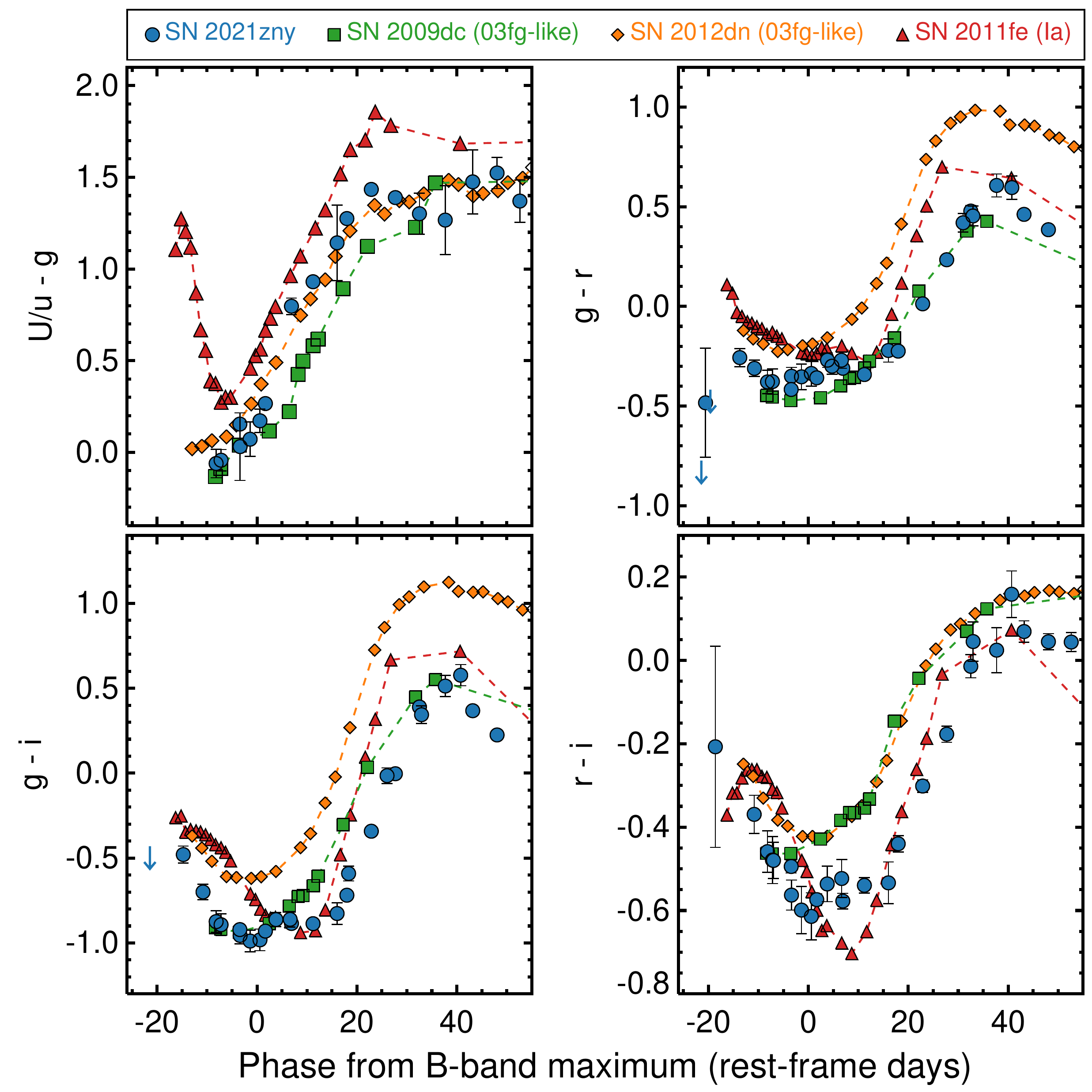}
\caption{The optical (\textit{$u-g$}, \textit{$g-r$}, \textit{$g-i$} and \textit{$r-i$}) colours of SN\,2021zny (blue circles), SN 2009dc (green squares), SN 2012dn (orange diamonds) and SN 2011fe (red upward triangles).}
\label{fig:colors}
\end{center}
\end{figure}

The optical colour evolution of SN\,2021zny is presented in Fig.~\ref{fig:colors}, compared with SNe 2009dc, 2012dn and 2011fe. At early times, SN\,2021zny exhibits considerably bluer colours than normal SNe Ia (probed by SN 2011fe), similar to SNe 2009dc and 2012dn. This discrepancy between the pre-maximum colour curves of SN\,2021zny and normal SNe\,Ia is particularly obvious in the bluer photometric bands, as can be seen in \textit{$u-g$}, which is a direct consequence of the small amount of line blanketing and the weak \ion{Ca}{ii} H\&K absorption features (Fig.~\ref{fig:spec_series}). Finally, another characteristic of the 03fg-like SNe Ia that separates them from normal SNe Ia is the \textit{$r-i$} colour evolution from peak up to about +30 d. The distinctive secondary maxima in these photometric bands and at these phases makes normal SNe Ia show a particularly sharp blue colour evolution, while 03fg-like SNe Ia appear considerably redder, with SN\,2021zny showing an intermediate \textit{$r-i$} colour evolution.

Of particular importance is the colour evolution during our earliest detections. At these epochs, which coincide with the flux excess detected in \textit{gri}, SN\,2021zny starts extremely blue (\textit{$g-r$}$\:\sim-0.5$), evolving to redder colours. SN\,2021zny is bluer than the earliest \textit{$g-r$} detection of SN 2011fe (at $\sim$ 2 days after explosion) by 0.6 mag. At later phases, SN\,2021zny settles to the usual `red-blue-red' colour evolution as seen in most thermonuclear SNe. This colour behaviour indicates that an external power source acts during the flux excess, with initially high temperatures (hence the blue early colour), that gradually subsides and allows the radioactive decay of $^{56}$Ni to dominate during the main light curve.

\subsection{The rising light curve} \label{sec:rising_lc}

As mentioned in Section~\ref{sec:discovery}, SN\,2021zny was discovered on UT 2021 September 22.37 by ZTF at $\textit{r}=19.33$~mag, however, forced photometry on ZTF images taken prior to discovery showed that the SN was also present in previous epochs. Adopting the distance and extinction on the line of sight from Section~\ref{sec:dist_extinction}, and our estimate of the time of maximum from Section~\ref{sec:phot_analysis}, SN\,2021zny is first detected (on UT 2021 September 19.50) at $\textit{g}=-15.32\pm0.34$~mag at $-21.38$ d from maximum ($\sim4.5$~mag fainter than its peak brightness), with simultaneous non-detections in $\textit{r}$- and $\textit{i}$-bands. Our last simultaneous non-detections in two photometric bands were at $-23.46$ d in $\textit{g}$- and $\textit{r}$-bands, with our last (i.e. a non-detection prior to our first $\textit{g}$-band detection) $\textit{i}$-band non-detection at $-26.34$ d. This indicates that SN\,2021zny exploded some time between $-23.46$ and $-21.38$ d from maximum, with potentially exceptionally high temperatures at the time after explosion, implied by its significant blue colours (Fig.~\ref{fig:colors}). Our strong constraint on the time of explosion is additionally corroborated by the {\it TESS} observations, as shown on the inset of Fig.~\ref{fig:phot}. We estimate the time of the first detection in the \textit{TESS}-band by following a similar approach as described in \citet{Dimitriadis2019ApJ}: we calculate the weighted-mean of the flux on a given time-window $x$, marking as a detection when $\mathrm{Flux}_{x}\ge3\times\sigma_{x-1}$, and iterate this procedure by reducing the width of the time-window. The final detection time and its uncertainty are then estimated as the mean and standard deviation of the recorded detection times. Using this method, we recover a time of first detection at UT 2021 September 19.28$\pm$0.16 ($\mathrm{MJD_{\textit{TESS},det}}=59476.28$), corresponding to $-21.60\pm0.15$ d with respect to \textit{B}-band maximum.

\begin{figure}
\includegraphics[width=\columnwidth]{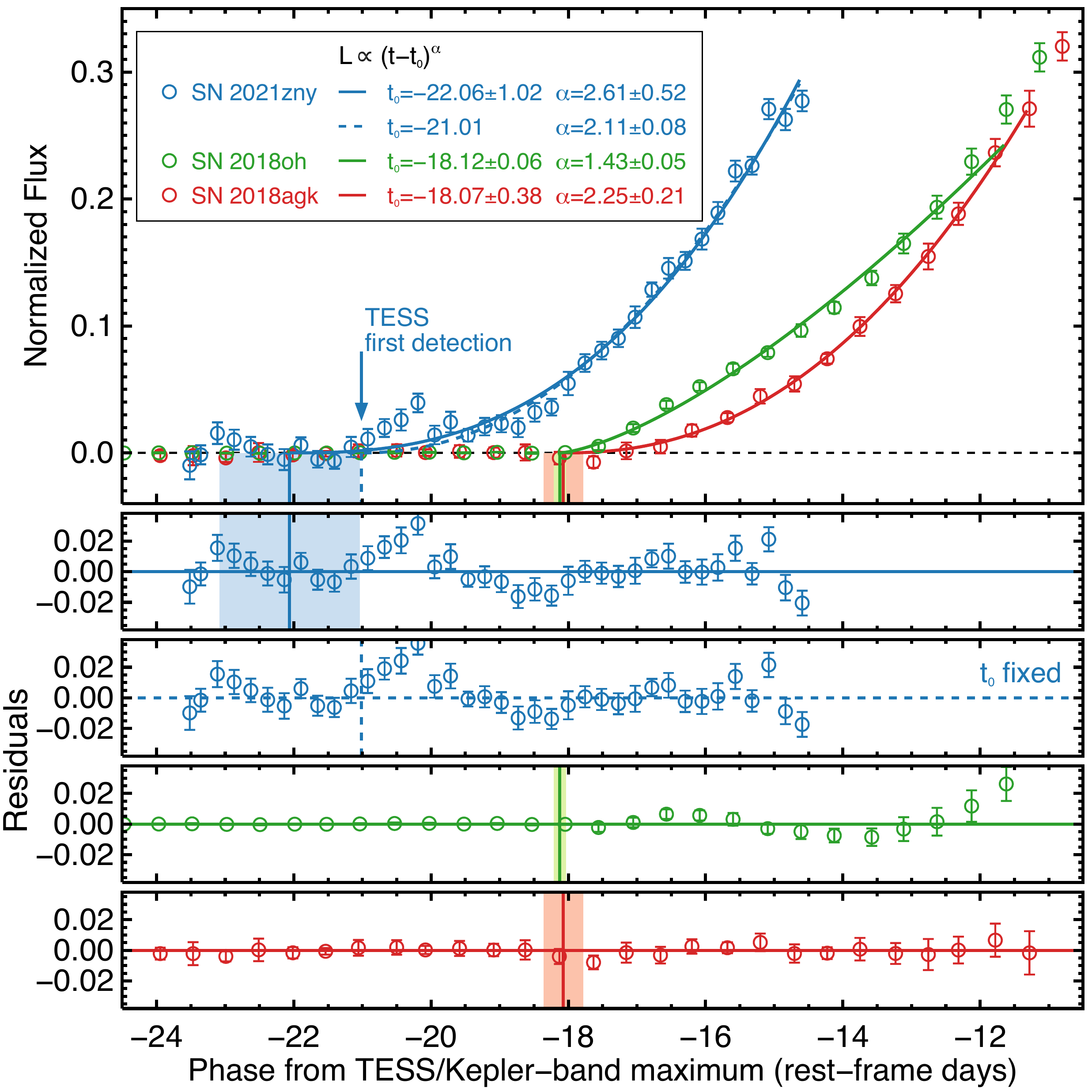}
\caption{Comparison of the SN\,2021zny {\it TESS} light curve, normalized to peak flux (open blue circles) with the {\it Kepler} light curves of SN\,2018oh (green) and SN\,2018agk (red), with respect to each photometric band's time of maximum. The time of {\it TESS} first detection is shown as a blue downward arrow. Power-law fits to the light curves, as described in the text, are additionally presented, with the fitted parameters shown in the legend and the residuals of the fits of each SN in the bottom panels.}
\label{fig:rise_fits}
\end{figure}

In Fig.~\ref{fig:rise_fits}, we present power-law rise fits (parameterised as $L\propto(t-t_{0})^{a}$, where $t_{0}$ is the time of first light and $a$ is the power-law index) to the \textit{TESS}-band light curve of SN\,2021zny. Due to the lack of \textit{TESS} observations after peak, an estimate of the \textit{TESS}-band time of maximum is uncertain, thus we use the \textit{r}-band time of maximum (UT 2021 October 10.86, $\mathrm{MJD_{\textit{r},peak}}=59497.86$) as the \textit{TESS}-band time of maximum. We compare our fits with similar fits of the \textit{Kepler}-band light curves of SN\,2018oh \citep{Dimitriadis2019ApJ}, a SN~Ia showing a prominent early flux excess, and SN\,2018agk \citep{Wang2021ApJ}, a SN~Ia with a smooth rising light curve, with their residuals additionally plotted. Shaded regions correspond to the 1-$\sigma$ uncertainty of our estimated time of first light. We note that, while the \textit{TESS}-band's transmission curve has a similar effective width as the \textit{Kepler}-band's ($\sim3,800$\:\AA), its central wavelength is approximately $1,500$\:\AA\: redder. Due to the gap of TESS observations from $-14.5$ until $-10.5$ d with respect to \textit{TESS}-band maximum, we restrict our time-range fit until the flux reaches 30 per cent of maximum (for all SNe), in contrast to the usual 40 per cent value that has been used in similar studies \citep[e.g. in][]{Olling15Natur}. We perform two separate fits for SN\,2021zny: the first is by keeping $t_{0}$ as a free parameter (shown with the solid blue line) and the second by fixing $t_{0}$ to the TESS time of first detection ($-21.01$ d, shown with the dashed blue line). For the first fit, we find $t_{0}=-22.06\pm1.02$ d and $\alpha=2.61\pm0.52$ ($\sim$1.05 days earlier than our first detection), while by fixing $t_{0}$, we find $\alpha=2.11\pm0.08$, with both values of the power-law index generally consistent with estimates in the literature \citep[e.g. see][]{Miller2020ApJ}. However, the residuals of both of our fits at the $-21$ to $-17.5$d phase-region resemble the characteristic `S-shape', as seen in SN\,2018oh, which is attributed to the presence of the early flux excess (although lasting significantly less), a behaviour not seen in the smooth rise of SN\,2018agk. While the significance of this deviation is lower than SN\,2018oh (due to lower fluxes), we conclude that the post-explosion light curve evolution of SN\,2021zny is in contrast with a single power-law rise, possibly due to the presence of an additional power source acting during a short period after the explosion.

\begin{figure}
\includegraphics[width=\columnwidth]{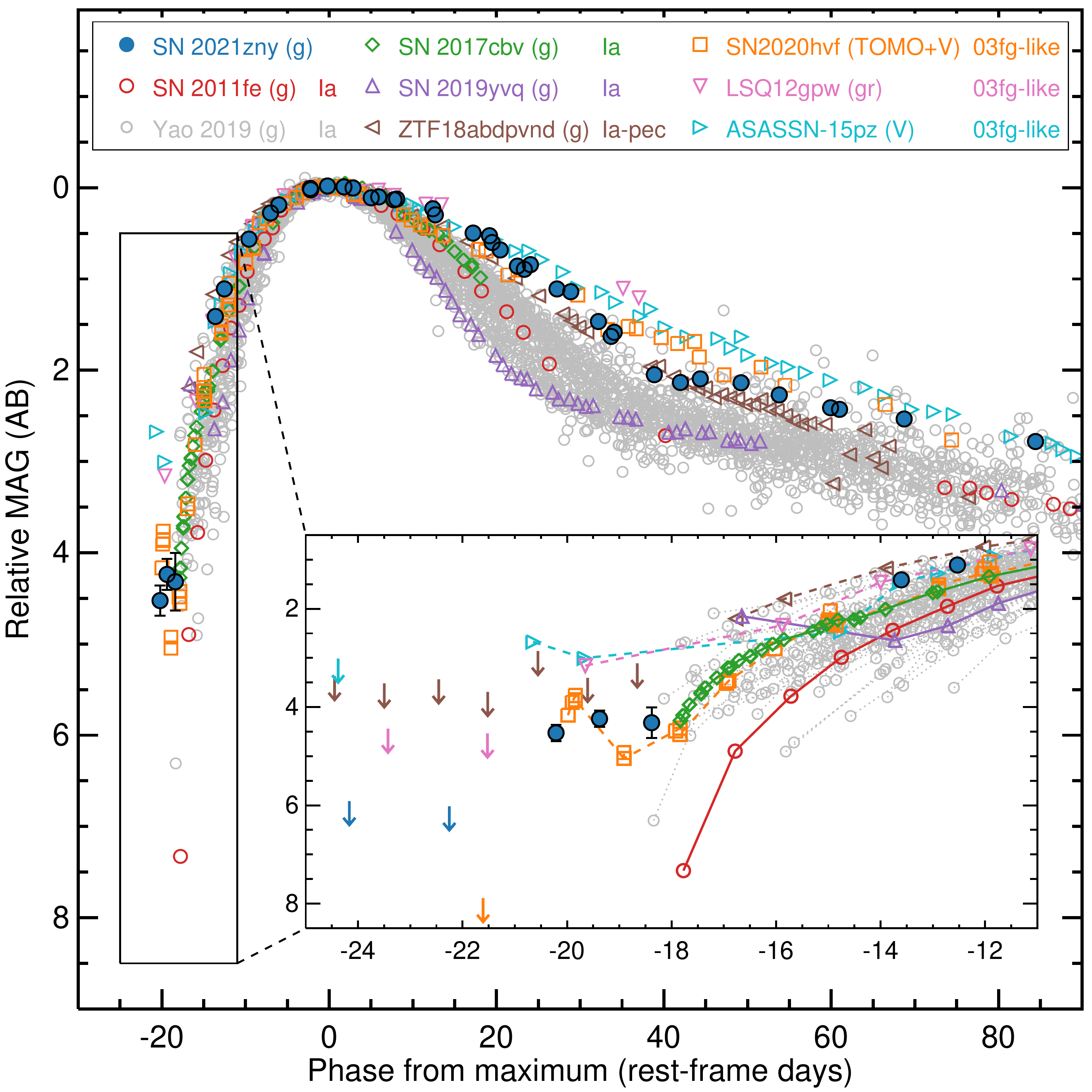}
\caption{Comparison of the k-corrected ZTF $\textit{g}$-band light curve of SN\,2021zny (full blue circles) with other SNe with early time data in similar photometric bands, as described in the legend. The light curves are normalised to their peak magnitude and presented with respect to each photometric band's time of maximum. In the inset, a zoom-in of the early epochs is shown, with downward arrows corresponding to non-detections of each SN.}
\label{fig:bump_comp}
\end{figure}

Fig.~\ref{fig:bump_comp} shows the k-corrected $\textit{g}$-band light curve of SN\,2021zny with respect to the $\textit{g}$-band time of maximum, which we estimate with low-order polynomial fits to be at UT 2021 October 10.25 ($\mathrm{MJD_{\textit{g},peak}}=59497.25$). We compare the light cure with the normal SN\,2011fe \citep{Firth2015MNRAS,Pereira13}, the slightly overluminous SN\,2017cbv \citep{Hosseinzadeh17ApJ}, the 03fg-like SNe 2020hvf \citep{Jiang2021ApJ}, LSQ12gpw \citep{Firth2015MNRAS} and ASASSN-15pz \citep{Chen2019ApJ} and the Ia-peculiar SN\,2019yvq \citep{Miller2020ApJ2}, in similar photometric bands. The inset panel provides a zoom-in of the light curves at the early epochs. We additionally plot the ZTF SNe Ia from \citealt{Yao2019ApJ} for $z<0.08$ (excluding the peculiar SNe Ia CSM and 02cx-like) and the only nearby 03fg-like SN Ia of that sample (ZTF18abdpvnd). K-corrections have been applied to the \citealt{Yao2019ApJ} sample and the relatively distant LSQ12gpw ($z=0.058$) and ASASSN-15pz ($z=0.015$).

The post-peak light curve evolution displays the established SN~Ia diversity, known as the width-luminosity relation \citep[WLR;][]{Phillips93}: brighter events, such as the 91T/99aa-like SNe Ia, decline slower than fainter events, such as the 86G/91bg-like ones. However, this diversity appears to extend at the early light curve evolution, for which we see smooth-rise events, (SN\,2011fe), strong long-lasting `bumps' (SN\,2017cbv) and `spikes' (SN\,2019yvq), and weaker short-lasting flux excesses (ASASSN-15pz, LSQ12gpw and SN\,2020hvf). Observationally, the early light curve of SN\,2021zny resembles that of other 03fg-like SNe Ia, showing a short ($\sim$2 days) and relatively weak flux excess as opposed to the long-lasting ($\sim$4--5 days) ones of SNe\,2017cbv and 2019yvq. 

For the 91T/99aa-like SN\,2017cbv and similar events, such as SNe\,2018oh \citep{Dimitriadis2019ApJ} and 2021aefx \citep{Hosseinzadeh2022ApJ}, and the Ia-peculiar SN\,2019yvq and similar events, such as iPTF14atg \citep{Cao2015Natur}, that show strong long-lasting flux excesses, proposed interpretations include the interaction of the ejecta with a non-degenerate companion, the presence of the radioactive $^{56}$Ni in the outer layers or the production of radioactive elements in the outer layers due to the nuclear burning in the He shell under a sub-$\mathrm{M_{ch}}$ double-detonation scenario. On the contrary, \citealt{Jiang2021ApJ} suggest that the short `flash' of SN\,2020hvf is more consistent with interaction of the SN ejecta, soon after explosion, with a confined and dense CSM, formed at the final evolution stage of the progenitor system. 

We note that not all 03fg-like SNe Ia may show this early `flash'; however, not many events of this subclass have been discovered early enough (the earliest detection of an 03fg-like SN Ia in the \citet{Ashall2021ApJ} sample is for SN\,2015M, at $-14.1$ d). Moreover, identifying a true flux excess is not trivial, as it requires early-time multi-wavelength observations with high cadence, limiting the promising candidates to very nearby events, but, at the same time, greatly reducing their potential numbers, due to the intrinsically low rate of 03fg-like SNe Ia. Nevertheless, for the events that were discovered early, such as SNe\,2021zny, 2020hvf and 2022ilv, their early flux excesses appear to originate from a different mechanism compared to events such as SNe\,2017cbv or 2019yvq, indicating a potential different explosion mechanism and/or progenitor binary configuration.

\subsection{Spectroscopic Properties} \label{sec:spec_analysis}

\begin{figure*}
\includegraphics[width=0.97\textwidth]{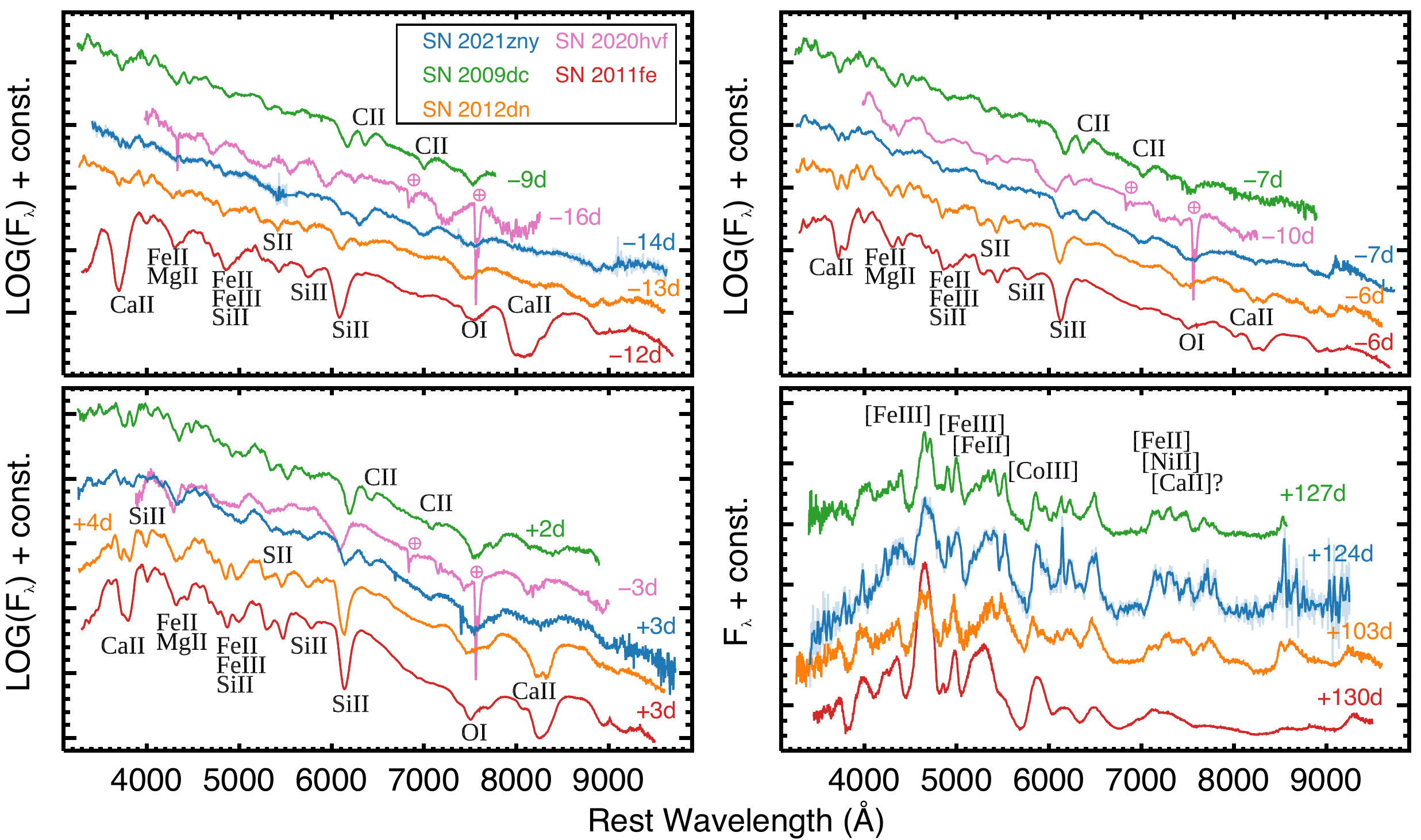}
\caption{Spectra of SN\,2021zny at $-$14 (top left), $-$7 (top right), $+$3 (bottom left) and $+$124 (bottom right) days from \textit{B}-band maximum are shown in blue, and are compared with spectra at similar phases of the 03fg-like SNe 2009dc (green), 2012dn (orange), 2020hvf (pink) and the normal SN Ia 2011fe (red), with their corresponding phases labelled. The spectra of SN\,2021zny have been deredshifted and corrected for extinction on the line of sight according to Section~\ref{sec:dist_extinction}, and for our comparison sample according to their relevant studies. The $-$14 and $+$124 d spectra of SN\,2021zny have been smoothed for presentation purposes, with light/dark blue corresponding to the raw/smoothed spectrum, respectively. All spectra are in flux density per unit wavelength, $F_{\lambda}$. Main absorption/emission features attributed to atomic species usually found in thermonuclear SNe in early/nebular phases are also marked.}
\label{fig:spec_comp}
\end{figure*}

Spectral comparisons of SN\,2021zny, in various phases of its evolution, with the 03fg-like SNe Ia 2009dc, 2012dn, 2020hvf and the normal SN~Ia 2011fe are presented in Fig.~\ref{fig:spec_comp}. 

The earliest spectrum of SN\,2021zny shows the general characteristics of the 03fg-like subclass:  particularly the blue (pseudo-) continuum (described reasonably well with a black body of $T\sim13,900$ K), the relatively strong absorption features of \ion{Si}{ii} $\lambda$6355, \ion{C}{ii} $\lambda\lambda$6580,7231 and \ion{O}{i} $\lambda$7774, and the extremely weak (or even absent) lines from other intermediate-mass elements (IMEs), such as \ion{S}{ii}, \ion{Ca}{ii} and \ion{Mg}{ii}. As the SN evolves toward maximum brightness, the usual IMEs and iron-group elements seen in thermonuclear SNe (e.g., \ion{Mg}{ii}, \ion{Ca}{ii}, \ion{S}{ii}, \ion{Fe}{ii} and \ion{Fe}{iii}) start to appear, but considerably weaker, compared to normal SNe Ia, in accordance with 03fg-like SNe Ia. We also note that the spectroscopic evolution of SN\,2020hvf seems to be different compared to other 03fg-like SNe Ia in our sample. For example, SN\,2020hvf displays unburned and synthesized material at remarkably higher velocities ($\sim10,500$ for SN\,2009dc and $\sim7,500$ \kms\ for SN\,2012dn at early times) and the \ion{C}{ii} $\lambda$6580 line appears relatively weak while other lines appear relatively strong (e.g. \ion{Si}{ii} $\lambda$5972). As SN\,2021zny evolves towards the nebular phase, its decreasing ejecta density and optical depth allow us to probe the inner layers of the SN, and forbidden emission lines from iron-group elements start to appear, particularly the forbidden lines [\ion{Fe}{iii}], [\ion{Fe}{ii}] and [\ion{Fe}{iii}].

\begin{figure}
\includegraphics[width=\columnwidth]{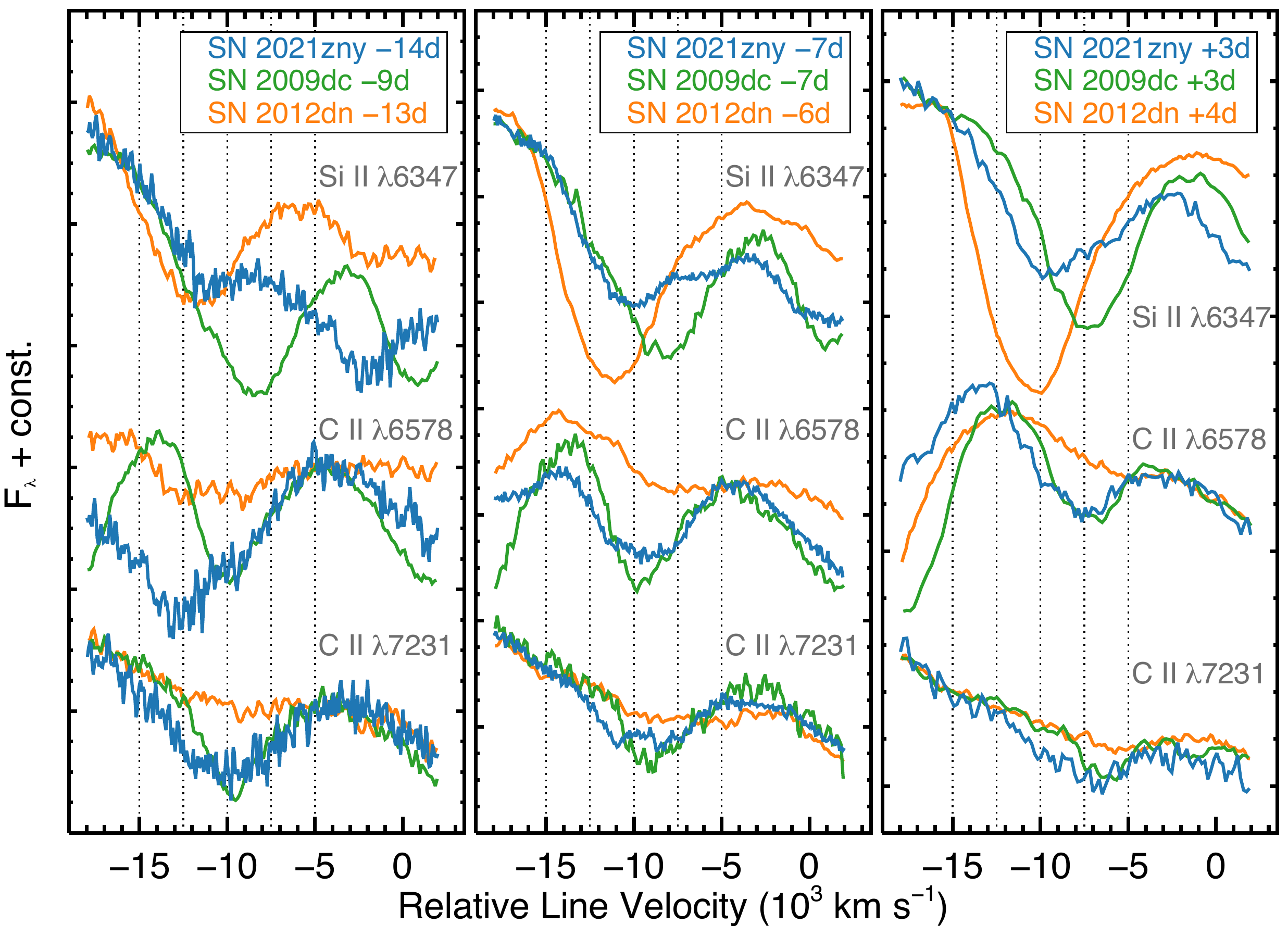}
\caption{Spectra of SN\,2021zny (blue), 2009dc (green) and 2012dn (orange) with respect to relative line velocities of \ion{Si}{ii} $\lambda$6355, \ion{C}{ii} $\lambda$6580 and \ion{C}{ii} $\lambda$7231, as labelled. The spectral epochs for each SN, with respect to \textit{B}-band time of maximum, is shown in the legends.}
\label{fig:spec_comp_vel}
\end{figure}

A noticeable characteristic of SN\,2021zny is the observed absorption line complex at $6,000$ -- $6,500$ \AA. As can be seen in Fig.~\ref{fig:spec_comp}, top left, the redder trough of the line complex (attributed to \ion{C}{ii} $\lambda$6580) is significantly stronger compared to the bluer one (attributed to \ion{Si}{ii} $\lambda$6355), indicating that at 2 weeks before maximum light, SN\,2021zny had far more unburned material above the photosphere, as compared to SNe\,2009dc and 2012dn. A close inspection to the spectra profiles reveals potentially two components for silicon and carbon for SN\,2021zny, as shown in Fig.~\ref{fig:spec_comp_vel}, where we show the spectra in velocity space with respect to the rest wavelengths of \ion{Si}{ii} $\lambda$6355, \ion{C}{ii} $\lambda$6580 and \ion{C}{ii} $\lambda$7231. At $-14$ d, two carbon components at approximately $-12,500$ and $-9,000$ \kms\ can be seen, with potentially two silicon components at similar velocities. The identification is more clear at $-7$d, where two silicon components are visible at approximately $-11,000$ and $-6,000\:\mathrm{km\:s^{-1}}$, with potentially two carbon components at similar velocities, seen in the \ion{C}{ii} $\lambda$7231 region. Moreover, the \ion{C}{ii} $\lambda$6580 feature appears much broader than what is seen in SN\,2009dc (FWHM of approximately $6,000$ \kms\ compared to $4,000$ \kms\ in SN\,2009dc), with the characteristic flattening of the minimum of the P-Cygni profile, which indicates a blend of two components. While no definite conclusion on the nature of these putative two components can be made, it is obvious that significant unburned material at a wide velocity range (probed by \ion{C}{ii} $\lambda\lambda$6580, 7231) persistently remains up to $\sim2$ weeks after maximum.

\begin{figure}
\includegraphics[width=\columnwidth]{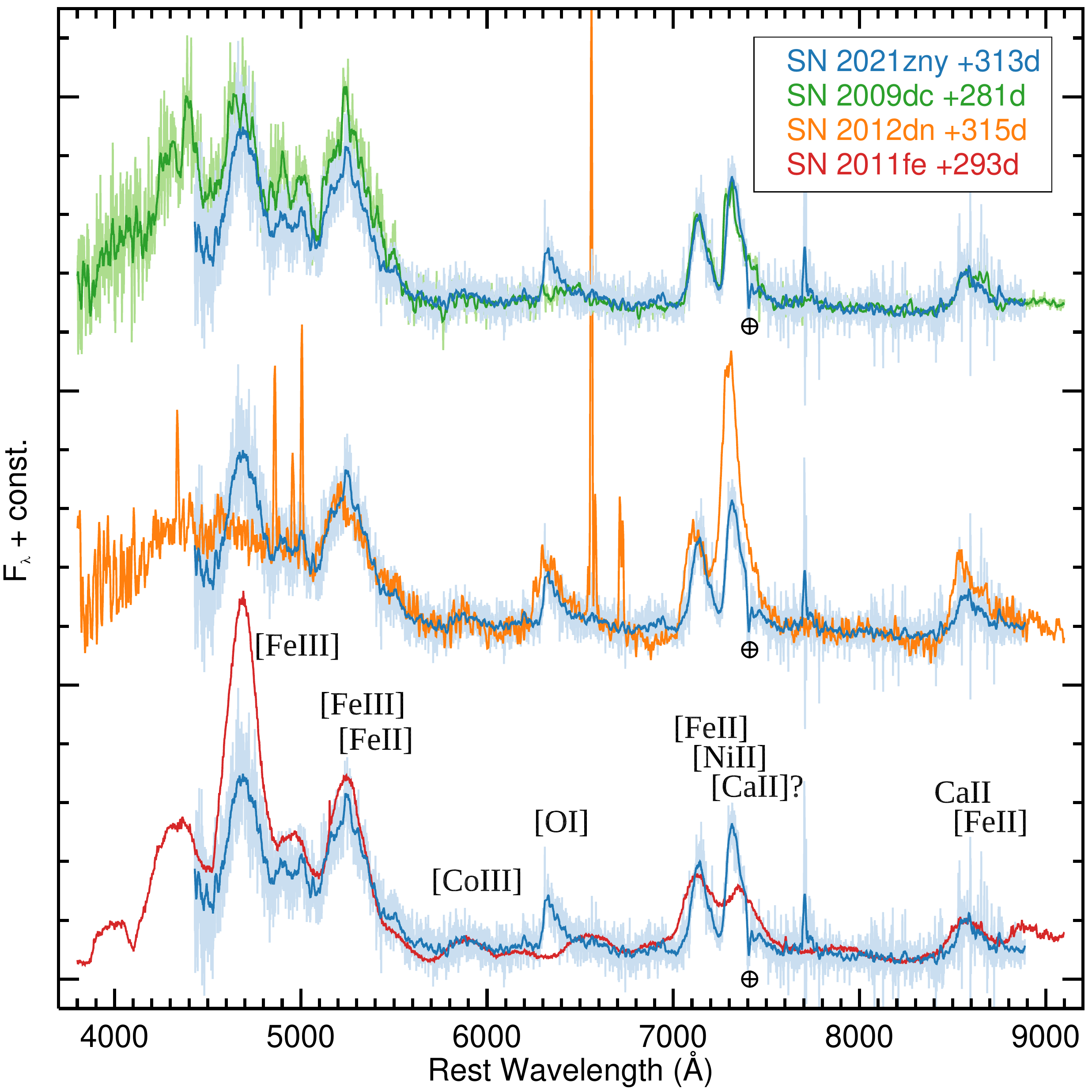}
\caption{Nebular time spectra of SN\,2021zny (blue), 2009dc (green), 2012dn (orange) and 2011fe (red), with the corresponding phases from \textit{B}-band maximum indicated in the legends. The spectra of SNe\,2021zny and 2009dc have been smoothed for presentation purposes, with light/dark colours corresponding to raw/smoothed spectra. All spectra have been normalized to the baseline flux measured around $8,000$ \AA. Various spectral features of normal and 03fg-like SNe Ia present in nebular epochs are marked. The strong and narrow emission lines in the SN\,2012dn spectrum originate from the host galaxy.}
\label{fig:spec_late}
\end{figure}

Fig.~\ref{fig:spec_late} shows our late-time ($313$ d from peak) spectrum of SN\,2021zny, compared with spectra of SNe 2009dc, 2012dn and 2011fe at similar phases. Our spectrum is remarkably similar to the one of SN 2009dc, showing the characteristic low [\ion{Fe}{iii}] to [\ion{Fe}{ii}] line ratio, attributed either to lower temperatures of the ejecta or higher ejecta densities, favouring enhanced recombination \citep[][]{Taubenberger2013MNRAS}. An additional argument for the low ionization state is the line complex at $7,000-7,500$ \AA. In normal SNe Ia, this feature is dominated by blends of [\ion{Fe}{ii}] and [\ion{Ni}{ii}], however, in SN\,2021zny (and in fact in most of 03fg-like events) two sharp emission peaks are observed. 

\begin{figure}
\includegraphics[width=\columnwidth]{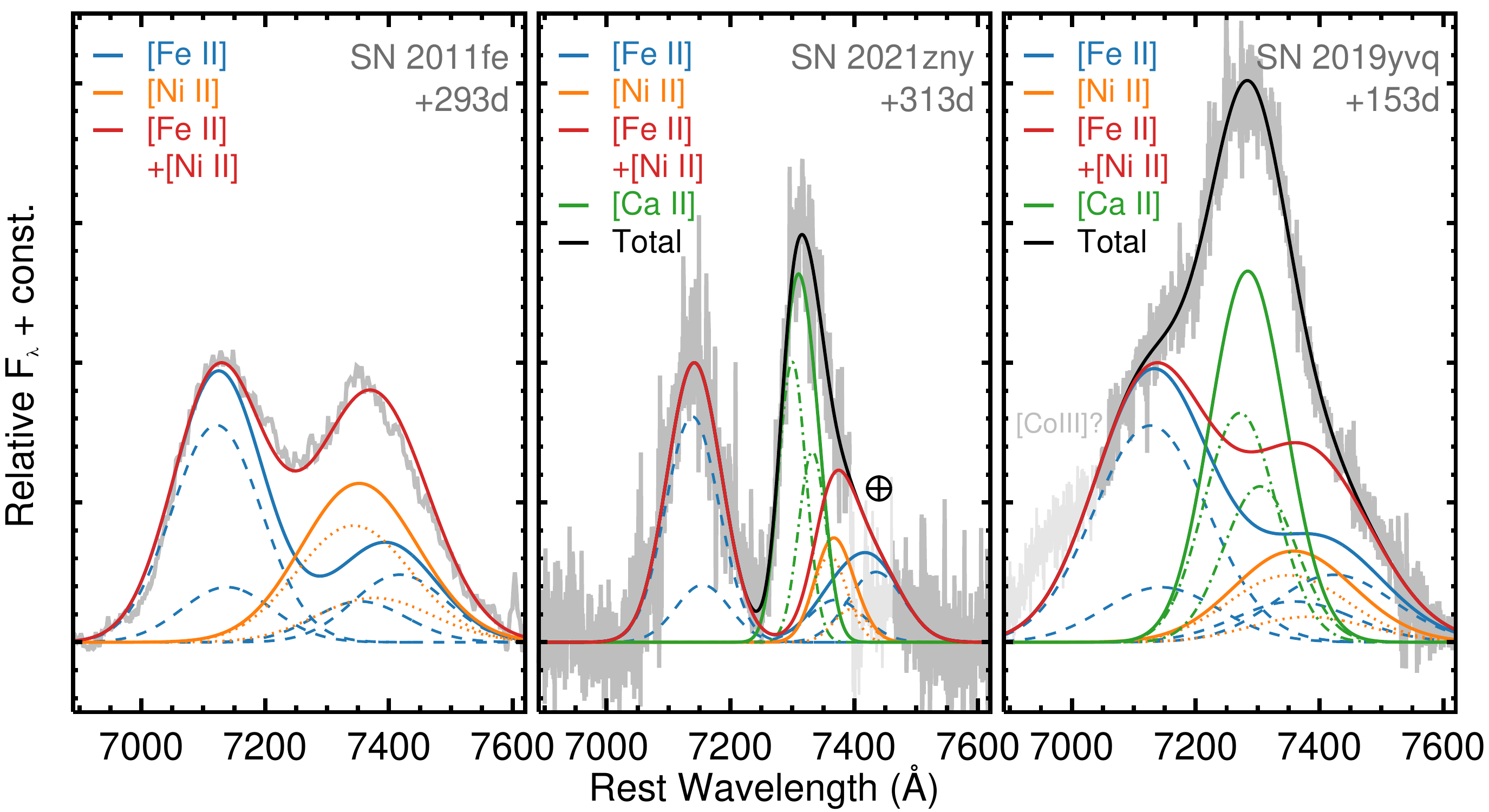}
\caption{The $7,000-7,500$ \AA\: spectral region of SN\,2011fe (left), SN\,2021zny (middle) and SN\,2019yvq (right), with each spectrum phase from \textit{B}-band maximum marked. Light grey regions correspond to spectral regions that are excluded from the fits (see text for details). Solid lines correspond to our best fits of the [\ion{Fe}{ii}]+[\ion{Ni}{ii}](+[\ion{Ca}{ii}]) model, as described in the text, with dashed, dotted and dashed-dotted lines corresponding to the individual lines of each atomic species. The spectra are normalised to the inferred peak of each [\ion{Fe}{ii}] feature ($\sim7,150$ \AA).}
\label{fig:spec_late_7000_7500}
\end{figure}

We attempt to model this emission feature as a sum of multiple Gaussian components for the [\ion{Fe}{ii}] (7155, 7172, 7388 and 7453 \AA) and [\ion{Ni}{ii}] (7378, 7412 \AA) blends, with the relative strengths A$_{i}$ of the individual lines for each atomic species tied as in \citet{Maguire2018MNRAS}, and a common velocity shift and FWHM. We find v$_{\mathrm{Fe}}=-735\pm35\:\mathrm{km\:s^{-1}}$, FWHM$_{\mathrm{Fe}}=4,175\pm85\:\mathrm{km\:s^{-1}}$, v$_{\mathrm{Ni}}=-2,710\pm25\:\mathrm{km\:s^{-1}}$ and FWHM$_{\mathrm{Ni}}=3,025\pm65\:\mathrm{km\:s^{-1}}$, with the relative strength ratio estimated as A$_{\mathrm{Ni}}$/A$_{\mathrm{Fe}}=1.36\pm0.03$, indicating a significant difference at the velocity shifts of iron and nickel, with the [\ion{Ni}{ii}] blend being stronger than [\ion{Fe}{ii}], as opposed to normal SNe Ia. We thus explore the possibility of a calcium contribution, and perform a fit by adding a double Gaussian for the [\ion{Ca}{ii}] $\lambda\lambda$7292, 7324 doublet. Due to the presence of the telluric A-band absorption in the spectrum at the expected location of [\ion{Ni}{ii}], we exclude this region from the fit and assume a common velocity shift for [\ion{Fe}{ii}] and [\ion{Ni}{ii}]. For consistency, we apply the same model to SN\,2011fe and to SN\,2019yvq, where its late time spectrum also shows strong evidence on the presence of calcium \citep{Siebert2020ApJ,Tucker2021ApJ}. We note that the spectrum of SN\,2019yvq was taken at $+153$ days, when the 77.2-day half-life cobalt radioactive decay still contributes; thus an emission feature at $\sim7,000$ \AA, which is attributed to [\ion{Co}{iii}], is excluded from the fit. Our results are shown in Fig.~\ref{fig:spec_late_7000_7500} and presented in Table~\ref{tab:neb_spec_measurements}.

\begin{table*}
\caption{Best-fit parameters of the $7,000-7,500$ \AA\: line complex measurements.}
\label{tab:neb_spec_measurements}
\begin{tabular}{lccccccc}
\hline
    Name & v$_{\mathrm{Fe}}$ & FWHM$_{\mathrm{Fe}}$ & v$_{\mathrm{Ni}}$ & FWHM$_{\mathrm{Ni}}$ & A$_{\mathrm{Ni}}$/A$_{\mathrm{Fe}}$ & v$_{\mathrm{Ca}}$ & FWHM$_{\mathrm{Ca}}$ \\
      & ($\mathrm{km\:s^{-1}}$) & ($\mathrm{km\:s^{-1}}$) & ($\mathrm{km\:s^{-1}}$) & ($\mathrm{km\:s^{-1}}$) &  & ($\mathrm{km\:s^{-1}}$) & ($\mathrm{km\:s^{-1}}$) \\
    \hline
SN\,2021zny$^{a}$ & $-735\pm35$ & $4,175\pm85$ & $-2,710\pm25$ & $3,025\pm65$ & $1.36\pm0.03$ & -- & -- \\
SN\,2021zny$^{b}$ & $-705\pm35$ & $4,270\pm85$ & $-705\:(\equiv$v$_{\mathrm{Fe}})$ & $2,650\pm610$ & $0.38\pm0.05$ & $345\pm35$ & $2,155\pm100$ \\
SN\,2011fe & $-1,420\pm70$ & $7,180\pm120$ & $-1,420\:(\equiv$v$_{\mathrm{Fe}})$ & $8,890\pm370$ & $0.54\pm0.01$ & -- & -- \\
SN\,2019yvq & $-1,170\pm40$ & $9,115\pm435$ & $-1,170\:(\equiv$v$_{\mathrm{Fe}})$ & $9,160\pm1,960$ & $0.31\pm0.02$ & $-830\pm40$ & $5,670\pm150$ \\
    \hline
\multicolumn{8}{l}{$^a$ No calcium component.}\\
\multicolumn{8}{l}{$^b$ Assuming a common velocity offset for [\ion{Fe}{ii}] and [\ion{Ni}{ii}].}\\
\end{tabular}
\end{table*}

The most striking characteristic of the late time spectrum of SN\,2021zny is the emission feature at $\sim6,300$ \AA, seen also in SN\,2012dn, marking the second ever detection of this feature in an 03fg-like SN Ia. This feature is routinely seen in core-collapse (and particularly stripped-envelope) SNe \citep[][]{Taubenberger2009MNRAS} and it has been identified as [\ion{O}{i}] $\lambda\lambda$6300, 6364, but has never been seen in normal SNe Ia. However, it has been observed in the low-luminosity/slowly-evolving 02es-like SN\,2010lp \citep[][]{Taubenberger2013ApJ} and iPTF14atg \citep[][]{Kromer2016MNRAS}, with the later, interestingly, displaying an early UV flux excess \citep[][]{Cao2015Natur}. 

We model the emission following \citet{Taubenberger2013ApJ}, and we use two Gaussian components to account for the doublet at 6300 and 6364 \AA\:. We adopt a relative intensity ratio of 3:1 (appropriate for the optically thin limit at nebular epochs) and the same relative velocity and FWHM. We find a velocity of $1,400\pm60\:\mathrm{km\:s^{-1}}$ and a FWHM of $2,615\pm120\:\mathrm{km\:s^{-1}}$ (similar to [\ion{Ca}{ii}]), with a total luminosity of $5.65(\pm0.21)\times10^{36}\mathrm{erg\:s^{-1}}$. While this line identification may be valid for low-luminosity SNe Ia (due to the low burning efficiency), it is difficult to reconcile with the 03fg-like ones \citep[][]{Taubenberger19}. If such an emission is indeed due to [\ion{O}{i}], it would imply the presence of substantial unburned material close the center of the ejecta, putting stringent constraints on the explosion model. Nevertheless, the [\ion{Ca}{ii}]/[\ion{O}{i}] ratio of SN\,2021zny is estimated to be $\sim2.5$, slightly higher than the one of SN\,2010lp and on the low extremes of the Ca-rich transients \citep[][]{Prentice2022MNRAS}.

Finally, several theoretical models \citep[e.g.][]{Lundqvist2013MNRAS,Boty2018ApJ} predict that, under a single-degenerate scenario, stripped material from the companion will be swept up and once the ejecta become optically thin, this material will emit, producing strong and relatively narrow (FWHM $\sim1,000\:\mathrm{km\:s^{-1}}$) hydrogen and/or helium emission features. Our late-time spectrum shows a narrow (FWHM$=82\pm18\:\mathrm{km\:s^{-1}}$) H$\alpha$ emission line at the host galaxy's redshift, thus, we deduce that it originates from the SN host galaxy, while no \ion{He}{i} $\lambda\lambda$5875,6678 lines are seen. In turn, this H$\alpha$ non-detection allows us to place an upper limit on the amount of stripped material following the method applied in \citet{Dimitriadis2019ApJ2}. As the \citet{Boty2018ApJ} synthetic spectrum was generated for an epoch of 200 d after peak, we scale our $313$ d spectrum using our photometry to estimate the $\textit{g}$-band magnitude at $200$ d, since the spectral features of SNe Ia do not change significantly between those epochs. We calculate an observed apparent magnitude of $\textit{g}_{200}=20.48\pm0.5$ mag, and after correcting for MW and host galaxy extinction and adopting the distance to CGCG 438-018, we infer an H$\alpha$ luminosity upper limit of $<2.6\times10^{37}\:\mathrm{erg\:s^{-1}}$, which corresponds to a hydrogen mass limit of $<6.4\times10^{-4}\:\mathrm{M_{\odot}}$. 

\subsection{The bolometric light curve} \label{sec:bol_lc}

\begin{figure}
\includegraphics[width=\columnwidth]{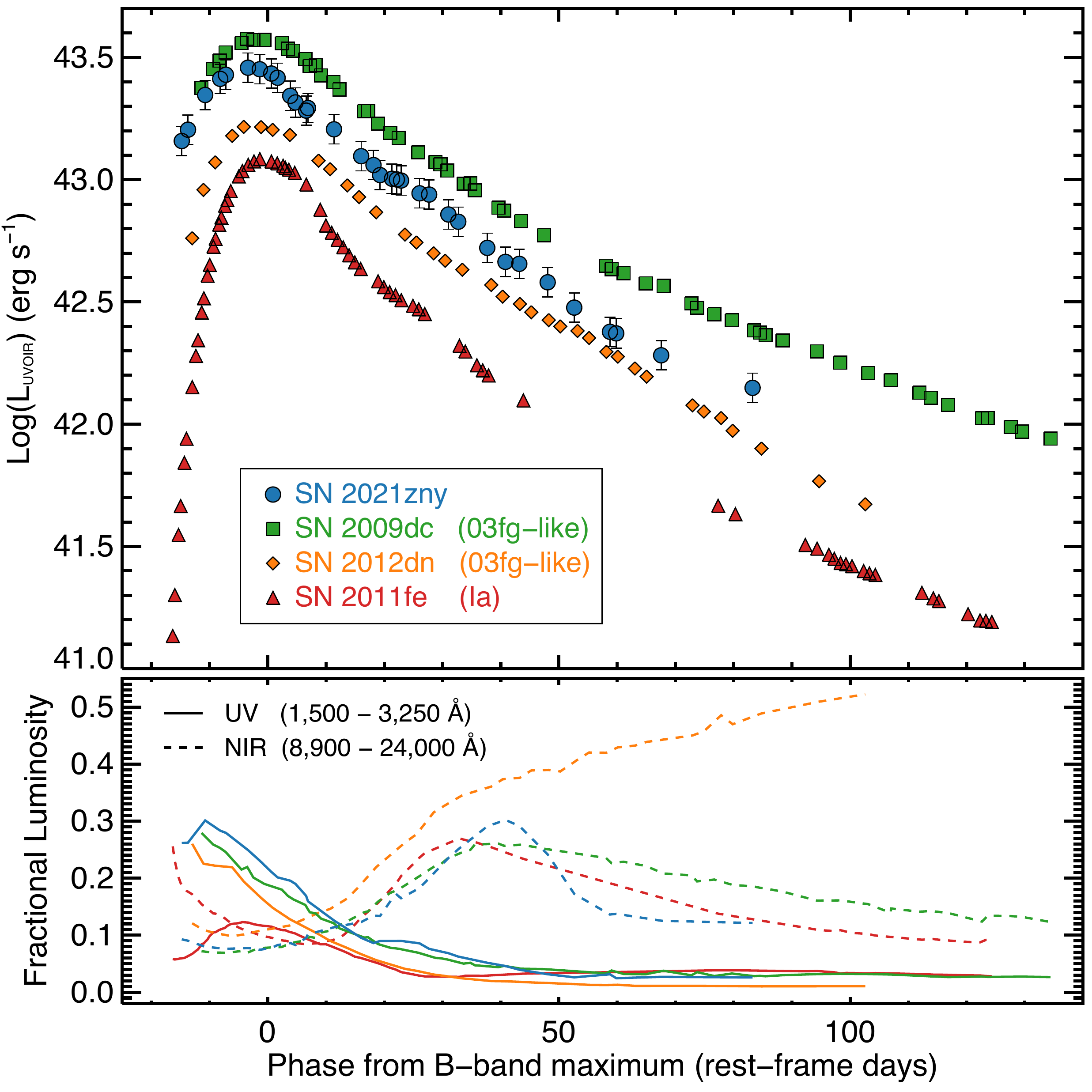}
\caption{The UVOIR (1,500 - 24,000 \AA) pseudo-bolometric light curve of SN\,2021zny, constructed as described in the text, is presented in blue circles. Similarly constructed light curves of SNe\,2009dc, 2012dn and 2011fe are additionally shown. On the bottom panel, we show the ratio of the UV (1,500 - 3,250 \AA, solid lines) and NIR (8,900 - 24,000 \AA, dashed lines) to the UVOIR luminosity of each SN.}
\label{fig:bol_lc}
\end{figure}

We constructed the UVOIR pseudo-bolometric light curve for SN\,2021zny from our broadband UV/optical/NIR photometry as follows: Firstly, we correct our photometry for Milky Way and host galaxy extinction (see Section~\ref{sec:dist_extinction}) and convert the magnitudes to monochromatic fluxes. We then interpolate our fluxes with Gaussian processes at the observed epochs of the $\textit{g}$-band photometric measurements, and assume zero flux for the UV and NIR bands after 2021 November 24 and December 04 UT, respectively. The computed spectral energy distribution is integrated (using the trapezoidal rule) with respect to each photometric band's effective wavelength, assuming zero flux at the blue end of the $\textit{UVW2}$-band (1,500 \AA) and the red end of the $\textit{K}$-band (24,000 \AA). Finally, we converted the integrated flux to luminosity using the distance to SN\,2021zny from Section~\ref{sec:dist_extinction}, with the uncertainty in the luminosity dominated by the uncertainty in the distance. We note that, in our calculations, we excluded epochs prior to 2021 September 9 UT (when the early flux excess is detected), as only $\textit{gri}$ and $\textit{TESS}$-band observations were acquired, and extrapolating the UV and NIR light curves to those phases is highly uncertain. We used the same procedure for SNe\,2009dc, 2011fe and 2012dn and construct similar pseudo-bolometric light curves, using the published photometry, extinction estimates and distances. Our final bolometric light curves are shown in Fig.~\ref{fig:bol_lc} and presented in Table~\ref{tab:bol_lc}.

\begin{table}
\caption{The UVOIR pseudo-bolometric light curve of SN\,2021zny.}
\label{tab:bol_lc}
\begin{tabular}{ccc}
\hline
    Phase$^a$ & Luminosity & Luminosity error \\
     (Rest-frame Days) & ($10^{43}\:\mathrm{erg\:s^{-1}}$) & ($10^{43}\:\mathrm{erg\:s^{-1}}$) \\
    \hline
-14.78 & 1.44 & 0.20 \\
-13.70 & 1.60 & 0.22 \\
-10.76 & 2.22 & 0.31 \\
-8.18 & 2.59 & 0.36 \\
-7.20 & 2.69 & 0.37 \\
    \hline
\multicolumn{3}{l}{$^a$ Relative to $B$-band maximum (MJD 59498.46).}\\
\multicolumn{3}{l}{This table is available in its entirety in machine-readable form.}\\
\end{tabular}
\end{table}

A common characteristic of the UVOIR bolometric light curves of the 03fg-like SNe Ia is the shift of the time of their peak luminosities, compared to their \textit{B}-band maximum. For the normal Type Ia SN\,2011fe, the peak luminosity occurs $\sim0.4$ d earlier than its \textit{B}-band maximum, while for SNe\,2021zny, 2009dc and 2012dn, we measure $\sim3.2, 1.4$ and $1.8$ d. This can be explained by the significant UV contribution to the bolometric light curve at early times, as illustrated by the bottom panel of Fig.~\ref{fig:bol_lc}. At the earliest epochs (15-10 d before \textit{B}-band maximum) of the three 03fg-like SNe Ia, the UV contribution is $\sim$25-30 per cent while for SN\,2011fe it is $\sim$6 per cent, with this reduced UV luminosity in normal Type Ia SNe attributed to increased line blanketing from iron peak elements, due to the increased UV opacity near the photosphere \citep{Mazzali2000AA}. For the case of the increased UV luminosity of 03fg-like SNe Ia, possible interpretations include differences in metallicity and/or the outer density structure of the ejecta \citep{Mazzali2014MNRAS}, although these differences cannot reproduce the observed diversity in the UV colours of SNe Ia \citep{Brown2015ApJ}. However, an additional UV-bright power source, apart from the radioactive decay of $^{56}$Ni, contributing substantially at earlier times could be a natural explanation, such as the interaction of the ejecta with surrounding CSM.

As mentioned above, we chose to exclude from our estimate of SN\,2021zny's bolometric light curve the extremely early epochs ($-$21 to $-$18 d with respect to \textit{B}-band maximum) due to limited photometric coverage, particularly in the UV. However, motivated by the sizable UV contribution, we attempt to characterize this early luminosity by fitting a black body to the early $\textit{gri}$-band data. We recover temperatures of 29,400 , 21,000 and 13,000 K at $20.6$, $19.6$ and $18.6$ d from \textit{B}-band maximum. While undoubtedly some emission from the $^{56}$Ni radioactive decay is present at those epochs, it is potentially subdominant: a fit to the early Lick/Kast spectrum of SN\,2011fe \citep{Nugent11}, taken $\sim1.5$ d after explosion ($\sim$16.3 d from \textit{B}-band maximum) gives a temperature of $7,700$ K, which corresponds to a difference of $\sim17,500$ K at this epoch.

Focusing on the UVOIR bolometric light curve at peak, we can estimate the $^{56}$Ni and ejecta masses of SN\,2021zny, \textit{assuming} that the near-peak luminosity evolution of the SN is powered solely by the radioactive decay of $^{56}$Ni and $^{56}$Co. We use the analytic model of \citet{Khatami2019ApJ}, an updated version of the classic \citet{Arnett82} model, that generates results in agreement with numerical radiation transport calculations. We use $\beta=1.6$, appropriate for SNe Ia (see their Table 2), a peak luminosity of $2.81(\pm0.05)\times10^{43}\:\mathrm{erg\:s^{-1}}$ and a rise time of $18.4\pm0.5$ d (corresponding to the rest-frame time from the $\textit{TESS}$-band first detection up to the peak of the UVOIR bolometric light curve), estimated with Gaussian processes fitting. We further assume a constant opacity of $\kappa=0.1\:\mathrm{cm^{2}\:g^{-1}}$ \citep[appropriate for iron-group element dominated ejecta, see][]{Pinto2000ApJ,Piro13ApJ} and an ejecta velocity of $v_{\mathrm{ej}}=10,000\:\mathrm{km\:s^{-1}}$, obtained from the absorption minimum of \ion{Si}{ii} $\lambda$6355 near maximum (see Fig.~\ref{fig:spec_comp_vel}). We obtain $M_{\mathrm{^{56}Ni}}=1.37\pm0.02\:\mathrm{M_{\odot}}$ and $M_{\mathrm{ej}}=1.60\pm0.10\:\mathrm{M_{\odot}}$. Adopting the same method, we estimate for the 03fg-like SNe\,2009dc and 2012dn $M_{\mathrm{^{56}Ni}}=1.79\pm0.02\:\mathrm{M_{\odot}}$ and $M_{\mathrm{ej}}=2.32\pm0.22\:\mathrm{M_{\odot}}$, and $M_{\mathrm{^{56}Ni}}=0.79\pm0.01\:\mathrm{M_{\odot}}$ and $M_{\mathrm{ej}}=1.73\pm0.11\:\mathrm{M_{\odot}}$, respectively, while for the normal SN\,2011fe we estimate $M_{\mathrm{^{56}Ni}}=0.57\pm0.01\:\mathrm{M_{\odot}}$ and $M_{\mathrm{ej}}=1.40\pm0.02\:\mathrm{M_{\odot}}$. Our mass estimates place SN\,2021zny in the 03fg-like regime, indicating a possible super-$\mathrm{M_{Ch}}$ mass origin for its progenitor (particularly in terms of its ejecta mass), with an enormous $^{56}$Ni synthesised mass, which is difficult to reconcile with current explosion models. We defer to Section~\ref{sec:discussion} for a comprehensive discussion.

Finally, at later times, SN\,2021zny generally shows a similar decline rate to SN\,2009dc. Our bolometric light curve coverage ends at $\sim90$ d from \textit{B}-band maximum and the increased fading seen in other 03fg-like events \citep[e.g. at $60$ d for SN\,2012dn, $>180$ d for SN\,2009dc and $>110$ d for SN\,2020esm;][]{Dimitriadis2022ApJ} has not occurred. We also cannot assess whether it happened at later times. For the case of SN\,2012dn, this rapid change in the decline rate was associated with a simultaneous decrease in the optical and increase in the NIR flux (see bottom panel of Fig.~\ref{fig:bol_lc}), something we do not see in SN\,2021zny up to these epochs (although we caution that the NIR flux at these epochs is poorly constrained and is based on extrapolation). However, the similarity of SN\,2021zny's light curves with that of SN\,2009dc (particularly in the NIR) indicates that this, if it happened, may have occurred at later stages of its evolution.

\section{Discussion} \label{sec:discussion}

In this section, we discuss the results of our analysis in the context of proposed progenitor systems of the 03fg-like SN Ia subclass. We summarise that a viable progenitor model for SN\,2021zny must address the early flux excess, the increased early UV luminosity, the high luminosity at peak relative to its decline rate, the persistent and strong carbon features (a proxy of unburned material), the absence of narrow features (particularly hydrogen) in the spectra and the presence of oxygen, and potentially calcium, emission at nebular epochs.

\begin{figure*}
\includegraphics[width=0.97\textwidth]{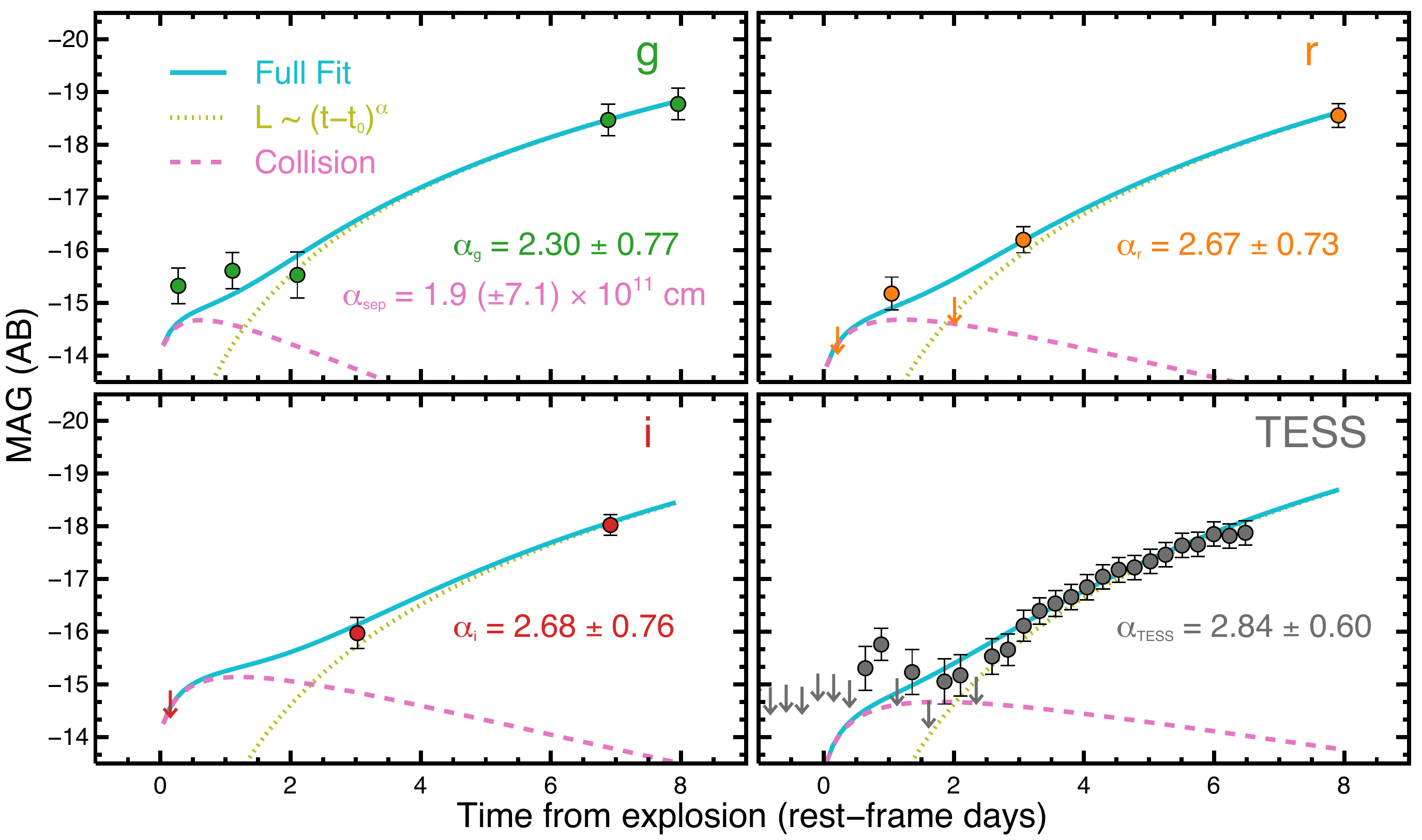}
\caption{Early ZTF $\textit{g}$- (top left), $\textit{r}$- (top right), $\textit{i}$- (bottom left) and $\textit{TESS}$-band (bottom right) absolute magnitude light curves of SN\,2021zny, with our \citet{Kasen2010ApJ} model fits (Section~\ref{sec:discussion}) presented as solid lines. The collision and power-law model components are shown as dashed and dotted lines, respectively. The inferred parameters of our fits are labelled.}
\label{fig:my_kasen_fit}
\end{figure*}

As discussed in Section~\ref{sec:rising_lc}, the early flux excess seen in SN\,2021zny's $\textit{gri}$ and $\textit{TESS}$-band light curves mostly resembles the fast early flash seen in SNe 2020hvf and 2022ilv, and appears morphologically different from other flux excesses, such as the ones of SNe\,2017cbv and 2018oh. The relatively short timescale of the flux excess generally excludes the shock interaction between the supernova ejecta and a non-degenerate binary companion \citep{Kasen2010ApJ}, as for most favourable viewing angles the duration of the excess is 3--6 d, with its shape appearing more like a `bump' than a `spike' \citep[see Fig.~2 of][]{Jiang2018ApJ}. Moreover, the stringent constraint on the stripped material from the non-degenerate companion additionally disfavours this scenario \citep[however, see the discussion in][]{Dimitriadis2019ApJ2}. Nevertheless, we attempt a fit of the early light curves with a model which is a combination of two luminosity sources that power the light curve:  i) a collision-powered luminosity prescribed as in \citet{Kasen2010ApJ} and ii) a SN-powered luminosity (i.e.~the luminosity due to the radioactive \nick\ decay in the ejecta), for which we use a simple power-law, $L_{x}=C_{x}(t-t_{0,x})^{\alpha_{x}}$ \citep[see also][for an identical approach]{Ni2022arXiv}. We simultaneously fit the $\textit{gri}$ and $\textit{TESS}$-band light curves, with the free parameters of our model being the binary separation $a_{\mathrm{sep}}$ and the parameters describing the power-law rise $C_{g,r,i,TESS}$ and $\alpha_{g,r,i,TESS}$. The best-fitting results are shown in Fig.~\ref{fig:my_kasen_fit}.

We find a seperation at the moment of the SN explosion of $1.9(\pm7.1)\times10^{11}\:\mathrm{cm}$ and power-law indexes for each photometric band of $\alpha_{g}=2.30\pm0.77$, $\alpha_{r}=2.67\pm0.73$, $\alpha_{i}=2.68\pm0.76$ and $\alpha_{TESS}=2.84\pm0.60$, with the inferred power-law indexes lying on the extreme end of their distributions \citep[particularly for the redder bands,][]{Olling15Natur,Miller2020ApJ}. Assuming the companion fills its Roche lobe, this separation is consistent with a low-mass main-sequence (MS) or (marginally) a helium-star \citep{Liu2015MNRAS}. The MS companion is disfavoured due to the non-detection of $H\alpha$ at the nebular spectrum. For the case of a helium-star companion, the absence of hydrogen and the presence of [\ion{Ca}{ii}] and [\ion{O}{i}] can be naturally explained \citep{Lundqvist2013MNRAS}, as oxygen and calcium are better coolants compared to helium. However, the velocity range of calcium and oxygen in the helium-star companion interaction scenario is expected to be at $<1,000\:\mathrm{km\,s^{-1}}$ \citep{Pan2010ApJ,Pan2012ApJ,Liu2013ApJ}, while in the case of SN\,2021zny we measure $2,340\pm100$ and $2,615\pm120\,\mathrm{km\:s^{-1}}$ (see Section~\ref{sec:spec_analysis} and Table~\ref{tab:neb_spec_measurements}). More importantly, a single-degenerate scenario with a \citet{Kasen2010ApJ}-like interaction, while able to partially explain the early flux excess, is not able to reproduce most of the peak-time properties of SN\,2021zny (and 03fg-like SNe Ia in general), particularly the broad light curve in combination with the high luminosity, the observed blue colours, the low expansion velocities and the substantial amount of unburned material (see Section~\ref{sec:intro}).

We additionally investigate the possibility of varying $^{56}$Ni distributions, that can mimic nickel mixing to the outer layers of the ejecta, as a source of the early flux excess by comparing the early SN\,2021zny light curves with the models of \citet{Magee2020AA} and \citet{Magee2020AA2}. As pointed out by \citet{Magee2020AA}, extended nickel distributions are not able to reproduce `bumps' or `spikes', such as the one observed in the 03fg-like LSQ12gpw (see their table 2 and figure C.1). On the other hand, nickel shells at the outer edge of the ejecta can potentially introduce a flux excess, with \citet{Magee2020AA2} considering the light curves of SNe\,2017cbv and 2018oh. While the parameter space investigated in that study is rather small, we produce synthetic light curves of their models and we compared their early rise with SN\,2021zny. None of the models were able to match the observed SN\,2021zny flux excess, with the closest match being the SN\,2018oh model with a $0.01\:\mathrm{M_{\odot}}$ nickel shell and a width of $0.06\:\mathrm{M_{\odot}}$ in mass coordinates. However, this flux excess is still relatively long-lasting, while the rise time is significantly shorter and the resulting peak luminosity much lower. Finally, some sub-Chandrasekhar mass explosion models from massive C/O WDs with a thick helium-shell undergoing a double detonation can potentially match the high luminosity and slow decline of SN\,2021zny. However, their early flux excesses are less pronounced, their UV luminosity low and, crucially, no unburned material is present after explosion \citep{Polin2019ApJ}.

\begin{figure*}
\includegraphics[width=0.97\textwidth]{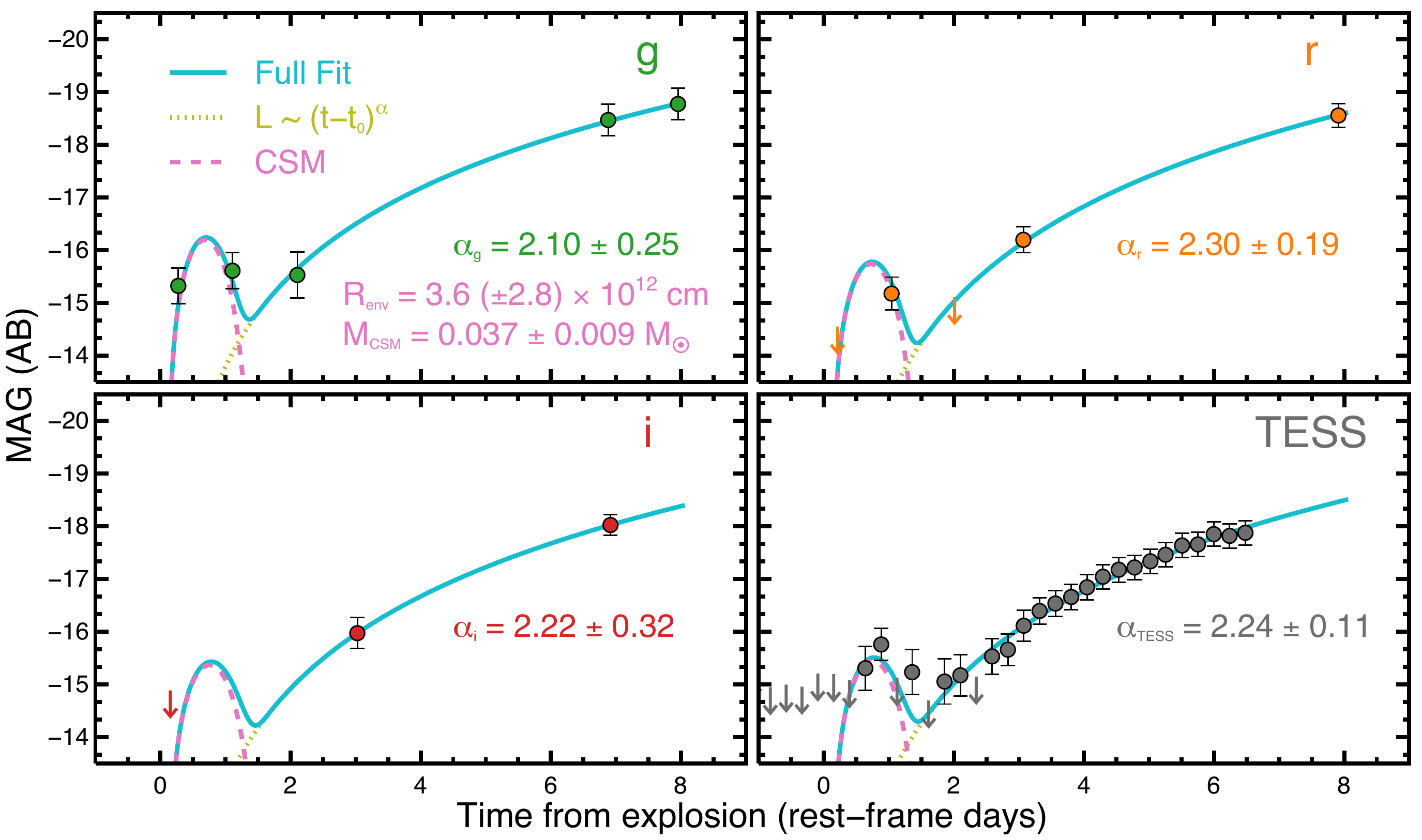}
\caption{Early ZTF $\textit{g}$- (top left), $\textit{r}$- (top right), $\textit{i}$- (bottom left) and $\textit{TESS}$-band (bottom right) absolute magnitude light curves of SN\,2021zny, with our CSM model fits (Section~\ref{sec:discussion}) presented as solid lines. The CSM and power-law model components are shown as dashed and dotted lines, respectively. The inferred parameters of our fits are labelled.}
\label{fig:my_csm_fit}
\end{figure*}

Motivated by the resemblance of SN\,2021zny to SNe 2020hvf and 2022ilv, we attempt to fit the early light curves of SN\,2021zny with a model similar to the one presented in \citet{Jiang2021ApJ}, which is a combination of two luminosity sources that power the light curve:  i) a CSM-powered luminosity assumed to originate from a spherically-symmetric envelope and ii) a SN-powered luminosity (i.e.~the luminosity due to the radioactive \nick\ decay in the ejecta). For the CSM-powered luminosity we use the prescription of \citet{Piro2015ApJ} and for the radioactive \nick\ decay, in photometric band $x$, we use a simple power-law, $L_{x}=C_{x}(t-t_{0,x})^{\alpha_{x}}$ \citep[see also][for an identical approach]{Ni2022arXiv}. The main difference of our model from the \citet{Jiang2021ApJ} one is the underlying SN light, for which they use a more realistic explosion model, developed with the radiation-hydrodynamic SuperNova Explosion Code \citep[SNEC;][]{Morozova2015ApJ}. We assume an electron-scattering opacity of $\kappa=0.2\:\mathrm{cm^{2}\:g^{-1}}$, appropriate for H-poor CSM \citep{Piro2015ApJ}, we fix the ejecta mass to the value obtained by our bolometric light curve estimates ($M_{\mathrm{ej}}=1.62\:\mathrm{M_{\odot}}$) and we break the degeneracy between the kinetic energy of the ejecta, $E_{\mathrm{ej}}$, and the ejecta velocity, $v_{\mathrm{ej}}$, by fixing the velocity to $v_{\mathrm{ej}}=10,000\:\mathrm{km\:s^{-1}}$ (as in Section~\ref{sec:bol_lc}). Finally, we assume that the time of the onset of the CSM interaction $t_{\mathrm{0,CSM}}$ coincides with the time of first light for each photometric band, i.~e. $t_{\mathrm{0,CSM}} \equiv t_{0,x}$. We simultaneously fit the $\textit{gri}$ and $\textit{TESS}$-band light curves, with the free parameters of our model being the mass $M_{\mathrm{env}}$ and radius $R_{\mathrm{env}}$, the time of the onset of the CSM interaction $t_{\mathrm{0,CSM}}$ and the parameters describing the power-law rise $C_{g,r,i,TESS}$ and $\alpha_{g,r,i,TESS}$. The best-fitting results are shown in Fig.~\ref{fig:my_csm_fit}.

Our final estimates for the envelope properties are $M_{\mathrm{env}}=0.003\pm0.001\:\mathrm{M_{\odot}}$ and $R_{\mathrm{env}}=3.6(\pm2.8)\times10^{12}$ cm. The total mass of the CSM $M_{\mathrm{CSM}}$ is then calculated assuming a $\rho \sim r^{-3}$ density distribution for the envelope, expected for post-merger CSM \citep[][]{Piro2016ApJ}, and adopting a progenitor WD radius of $3-6\times10^{8}$ cm \citep[][]{Piro2010ApJ} we find $M_{\mathrm{CSM}}=0.037\pm0.009\:\mathrm{M_{\odot}}$. The time of the onset of the CSM interaction is $0.06\pm0.05$ rest-frame days after $t_{\mathrm{det}}^{TESS}$ and the power-law indexes for each photometric band are $\alpha_{g}=2.10\pm0.25$, $\alpha_{r}=2.30\pm0.19$, $\alpha_{i}=2.22\pm0.32$ and $\alpha_{TESS}=2.24\pm0.11$. The inferred rise times and power-law indexes are well within estimates for overluminous SNe Ia \citep[][]{Olling15Natur,Miller2020ApJ}.
Removing the constraint of a common time of first light (but equal for every photometric band) and the onset of the interaction, we find $t_{\mathrm{0,CSM}}=0.08\pm0.05$ and $t_{0}=-0.41\pm0.98$ rest-frame days relative to $t_{\mathrm{det}}^{TESS}$, with $\alpha_{g}=2.32\pm0.60$, $\alpha_{r}=2.59\pm0.58$, $\alpha_{i}=2.43\pm0.61$ and $\alpha_{TESS}=2.52\pm0.62$, placing SN\,2021zny at the extreme end of the \citet{Miller2020ApJ} distribution, but still within limits. The subsequent CSM parameters are $M_{\mathrm{env}}=0.003\pm0.001\:\mathrm{M_{\odot}}$, $R_{\mathrm{env}}=3.9(\pm3.2)\times10^{12}$ cm and $M_{\mathrm{CSM}}=0.034\pm0.010\:\mathrm{M_{\odot}}$.

Our inferred CSM properties are similar to those of SN\,2020hvf and SN\,2022ilv, for which \citet{Jiang2021ApJ} and \citet{Srivastav2023ApJ} find $M_{\mathrm{CSM}}=0.01$ and $0.001$ \msun, respectively, with $R_{\mathrm{\mathrm{env}}}=1\times10^{13}$ cm. However, the epoch of SN\,2021zny's flux excess was observed in four photometric bands, including the dense $\textit{TESS}$-band coverage, as opposed to one band for SNe\,2020hvf and 2022ilv, allowing us to better constrain the black body temperature of the CSM-powered luminosity component and the time of explosion. Moreover, while our model includes several degeneracies and assumptions (particularly for $\kappa$, M$_{\mathrm{ej}}$ and v$_{\mathrm{ej}}$), it reasonably fits the observed light curves, sufficiently recovering the strength and timescale of the flux excess. Finally, we note that, as we do not use an explosion model for the SN-powered component but rather a power-law rise, therefore, possible effects that may alter the early SN light curve, such as nickel mixing \citep[e.g. see Fig. 13 and 14 of][]{Piro2016ApJ}, are not considered. Nevertheless, the three 03fg-like SNe Ia that were discovered extremely early show an early time light curve behaviour which is consistent with interaction of the SN ejecta with a dense shell of a low amount of CSM relatively close to the explosion site, indicating that this progenitor configuration might be common in the 03fg-like subclass.

A promising model that can explain the early flux excess is a merger of two C/O WDs, where the lower mass one is disrupted during the merging process. Hydrodynamical simulations of this binary configuration show that $\sim10^{-3}\:\mathrm{M_{\odot}}$ hydrogen and helium-free material from the disrupted WD can be ejected from the system, achieving escape velocities of $\sim2,000\:\mathrm{km\:s^{-1}}$ and resulting in material out to $10^{13}-10^{14}$ cm \citep{Raskin13}. The interaction of the SN ejecta with this C/O-rich CSM will result in additional UV/X-ray emission, and additional UV photons, as the shock-heated material cools. At the same time, the larger fraction of the disrupted WD will be quickly swept up by the ejecta, producing the strong and broad \ion{C}{ii} features \citep[][]{Raskin14} we observe in the early spectra of SN\,2021zny. As the unburned material from the disrupted WD forms an accretion disk around the exploding WD, strong orientation effects are expected, with equatorial viewing angles resulting in stronger \ion{C}{ii} features and lower ejecta velocities due to the increased deceleration. Several other parameters of this model can affect the observed properties, such as the mass ratio of the two WDs \citep{Dan2012MNRAS} and whether there is a delay between the disruption and the explosion \citep{Raskin13}.

\begin{figure}
\includegraphics[width=\columnwidth]{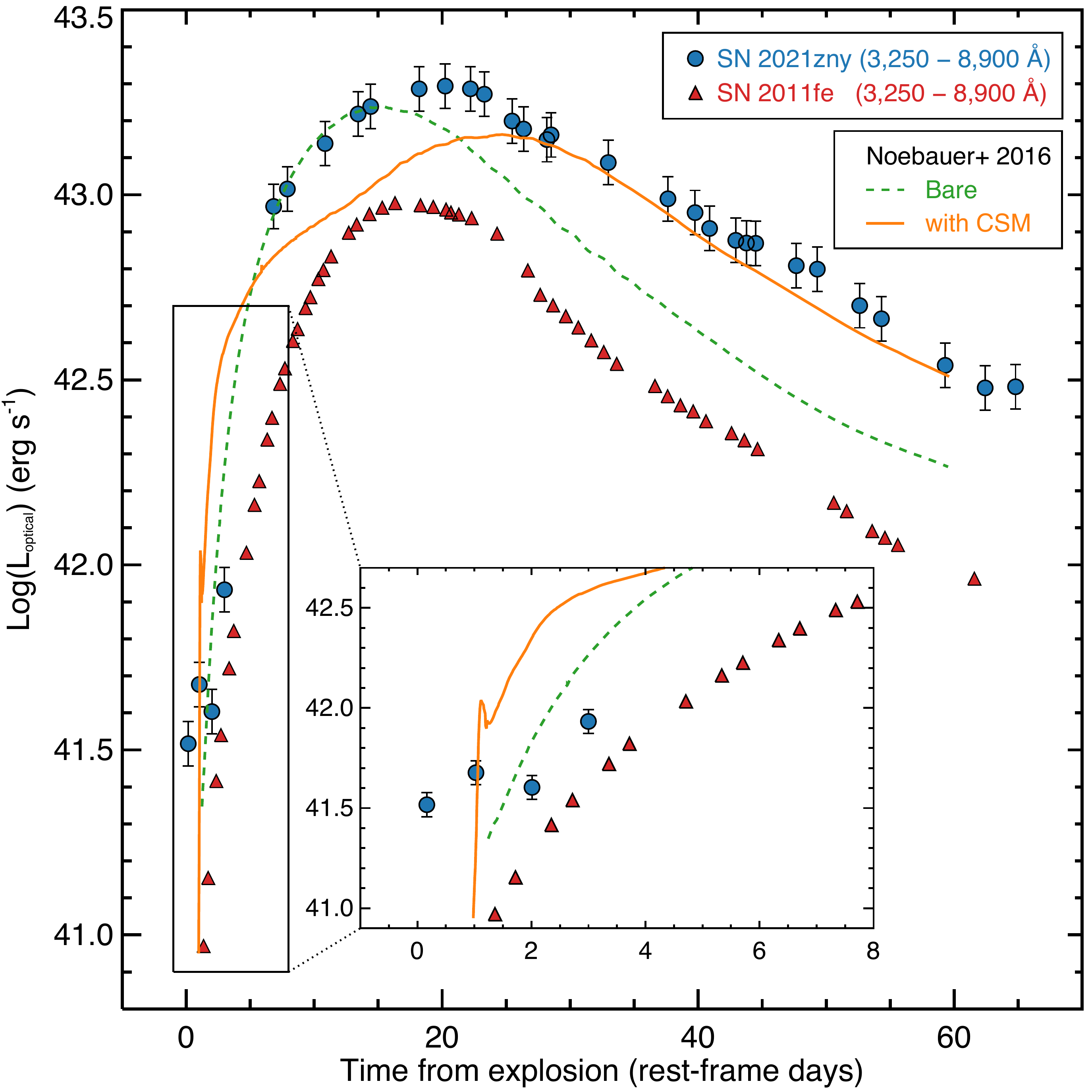}
\caption{The optical ($3,250-8,900$ \AA) pseudo-bolometric light curves of SNe\,2021zny (blue circles) and 2011fe (red upward triangles), compared with the bare ejecta (dashed green line) and the fiducial interaction (solid orange line) models of \citet{Noebauer16MNRAS}. In the inset, a zoom-in region of the early epochs is shown. The early flash of the interaction model, attributed to the high-energy radiation from the shock breaking out at the edge of the CSM \citep[see Fig. 5 from][]{Noebauer16MNRAS} coincides with the observed flux excess of SN\,2021zny.}
\label{fig:bol_noeb_comp}
\end{figure}

As already mentioned in \citet{Dimitriadis2022ApJ}, while these merger models can generally reproduce the spectroscopic properties of most 03fg-like SNe Ia, they fail in reproducing their bolometric light curves. \citet{Raskin14} present bolometric light curves for three merger configurations, for varying viewing angles, with the high total mass scenarios accurately predicting the peak luminosity but overestimating the width of the light curve (as the total ejecta mass is higher), while the situation is reversed for the lower total mass ones. We note that these models do not include the potential increased luminosity due to the interaction of the ejecta with the C/O-rich CSM. However, \citet{Noebauer16MNRAS} consider SN\,2009dc and present hydrodynamical and radiative transfer simulations of thermonuclear explosions of Chandrasekhar-mass WDs surrounded by relatively compact  ($R_{\mathrm{env}}\sim10^{14}\:\mathrm{cm}$) and massive ($M_{\mathrm{CSM}}\sim0.6\mathrm{M_{\odot}}$) C/O-rich CSM, which are able to generally reproduce the peak luminosity and the decline rate of 03fg-like events. Fig.~\ref{fig:bol_noeb_comp} shows the constructed optical ($3,250-8,900$ \AA) bolometric light curves of SNe\,2021zny (blue) and 2011fe (red), alongside the synthetic optical bolometric light curves from \citet{Noebauer16MNRAS} of a thermonuclear explosion of Chandrasekhar-mass WD with $1.4\:\mathrm{M_{\odot}}$ ejecta mass, producing $1.0\:\mathrm{M_{\odot}}$ of $^{56}$Ni (dashed green line) and the same model but embedded with $0.64\:\mathrm{M_{\odot}}$ C/O-rich material, extending to $1.3\times10^{14}$ cm (solid orange line). The `bare-ejecta' model fails to reproduce the width of the SN\,2021zny light curve and the time of maximum. Instead, it succeeds reproducing the general behaviour of the light curve of SN\,2011fe, but overestimates the peak luminosity, as SN 2011fe synthesized only $0.44\:\mathrm{M_{\odot}}$ of $^{56}$Ni \citep[][]{Pereira13}. The simulations involving the CSM roughly predicts the luminosity and the decline rate of SN\,2021zny at a few days after peak brightness and the overall width of the light curve, due to the increased trapping of the $\gamma$-rays from the additional CSM material and the increased density from the shock compression. The mismatch during the rising part of the light curve was addressed by increasing the content of non-carbon-oxygen material in the CSM, enhancing the opacity and the reprocessing efficiency \citep[see Fig. 7 and 8 from][]{Noebauer16MNRAS}. Nevertheless several parameters of this model can also alter the resulting light curves, with some of them already investigated by the authors, such as the mass and extent of the CSM. Moreover, in the inset we show the early model light curve, where a short-duration flash is seen, created from the reprocessing and shifting to the optical regime of the strong UV/X-ray luminosity, as a result of the shock breaking out at the edge of the CSM. We note that the earlier bump in SN\,2021zny is explained by interaction of the SN ejecta with a more compact CSM ($R_{\mathrm{env}}\sim10^{12}$ cm) soon after the explosion when the WD expanded to a radius of $\sim10^{11}\:\mathrm{cm}$ and detailed simulations of the ejecta-CSM interaction at early times are not performed in \citet{Noebauer16MNRAS}. 

We emphasise that both our early-time simple analytical model and the model addressing the near-peak evolution do not fully capture the physics of the ejecta-CSM interaction. Detailed numerical simulations, accounting for both the hydrodynamics and accurate radiative transfer calculations, are needed to thoroughly investigate the full parameter space of this model and simultaneously address all the phases of the phenomenon, which is not the scope of this paper. In a subsequent study, we will use the combined sample of SNe\,2020hvf, 2021zny and 2022ilv to explore several explosion models and CSM configurations, e.g. models with moderate $^{56}$Ni synthesized masses but increased additional luminosity from the CSM or asymmetries and multiple components of the CSM at more extended locations.

Our nebular observations of SN\,2021zny are crucial to constrain its progenitor system, since at these epochs the density of the ejecta is extremely low, unveiling the immediate environment of the SN explosion and its mechanism. The bolometric light curve up to +130 d from $B$-band maximum appears to follow a decline rate broadly consistent with the $^{56}$Co decay, with no sign of re-brightening (as would be expected with sustained CSM interaction) or accelerated decline (as seen in e.g. SN\,2012dn). At even later times, the absence in our nebular spectrum of hydrogen and helium emission features, either from stripped material from a potential non-degenerate companion or from H/He-rich CSM at larger distances, places strong constraints on the nature of the companion and the origin of the CSM. These observational characteristics are difficult to explain under the `core-degenerate' scenario \citep{Hoeflich96ApJ,Kashi11MNRAS,Hsiao20ApJ,Ashall2021ApJ}, as it predicts a late-time UV re-brightening, originating from the interaction of the ejecta with the AGB superwind, and narrow hydrogen/helium emission lines from the interaction of the ejecta with the AGB's envelope/wind. A possible solution to this problem is that the ejecta have not yet started to interact with the CSM, while the dense envelope prevents the $\gamma$-rays escaping and ionising this material. 

A double-degenerate merger scenario naturally explains the absence of H/He late time emission features, since the (low-mass) CSM is not H/He-rich, as it originates from the disrupted secondary C/O WD. Moreover, the detection of [\ion{O}{i}] $\lambda\lambda$6300, 6364 also favors a merger event, as this class of models are the only ones predicting the presence of unburned oxygen at low velocities, with the exception of turbulent pure deflagrations \citep{Kozma2005AA}. However, these models predict peculiar early time spectra and low luminosity explosions, clearly inconsistent with SN\,2021zny. The presence of oxygen at late times implies that calcium is also present (if these two elements are located in the same cooling region), as the [\ion{Ca}{ii}] $\lambda\lambda$7291, 7323 lines are $\sim100\times$ more efficient per atom in cooling than the [\ion{O}{i}] $\lambda\lambda$6300, 6364 lines \citep[][]{Fransson1989ApJ}, with the similar FWHM of the oxygen and calcium features of SN\,2021zny supporting this claim. If the red emission feature of the $7,000-7,500$ \AA\:line complex is indeed dominated by calcium, the remaining [\ion{Ni}{ii}] feature must be weak (relative to [\ion{Fe}{ii}]), implying a small amount of stable $^{58}$Ni synthesized in the explosion, since any nickel lines at nebular epochs cannot originate from the 6.1 days half-life radioactive $^{56}$Ni, while the iron lines are dominated by emission of the $^{56}$Co decay. We estimate the ratio of Ni to Fe abundance, following equation 1 from \citet{Maguire2018MNRAS}, to be $0.012\pm0.005$, decisively inconsistent with any Chandrasekhar-mass explosions ($0.17-0.20$) and consistent with some sub-Chandrasekhar mass explosions ($<0.15$), although the later are strongly disfavoured from the maximum light properties of SN\,2021zny. \citet{Maguire2018MNRAS} did not provide this ratio for any merger model, thus we use the abundances of the merger models of  \citet{Kromer2013ApJ}, \citet{Pakmor2010Natur} and \citet{Pakmor2012ApJ} (with WD masses of $0.9+0.76\:\mathrm{M_{\odot}}$, $0.9+0.9\:\mathrm{M_{\odot}}$ and $1.1+0.9\:\mathrm{M_{\odot}}$, respectively) obtained from the Heidelberg Supernova Model Archive\footnote{\url{https://hesma.h-its.org/}}, and we estimate ratios of $0.0002-0.0488$, consistent with SN\,2021zny \citep[however, see][for a discussion of this approach]{Blondin2022AA}.

While the properties of many 03fg-like SNe Ia (including SN\,2021zny) are in accordance with the merger scenario, some caveats still exist. The most important of them is the lack of observed asymmetries in the explosion \citep{Bulla16MNRAS}, probed by polarimetric observations \citep[][although these observations were performed after peak brightness]{Tanaka10ApJ,Cikota19MNRAS}. To this end, early time polarimetry of 03fg-like SNe Ia will prove extremely useful to disentangle between various proposed explosion models. Moreover, other model parameters, such as the mass ratio of the two WDs, the time delay between the disruption of the secondary WD during the merger and the final detonation of the primary WD and the location of the hotspot that causes the detonation (modifying the synthesized $^{56}$Ni mass) pose theoretical challenges \citep{Dan2011ApJ,Dan2012MNRAS,Raskin13}, but at the same time could account for the diversity within the 03fg-like subclass. Finally, a precise rate of the intrinsically rare 03fg-like SNe Ia has yet to be calculated, and as different progenitor models involve a diverse set of progenitor stars, a proper measurement of the rate, for which non-targeted all-sky surveys such as ZTF are essential, is highly encouraged.

\section{Conclusion} \label{sec:conclusion}

In this work, we presented extensive multi-wavelength photometric and spectroscopic observations of SN\,2021zny, the first 03fg-like event with a flux excess within the first $\sim1.5$ d after explosion, estimated at $-21.6$ d before maximum light, and, simultaneously, prominent [\ion{O}{i}] $\lambda\lambda$6300, 6364 emission lines at $313$ d after peak brightness. SN\,2021zny is the third member of its class, after SNe\,2020hvf and 2022ilv, with an early flash, and the first with a multi-wavelength photometric detection, allowing us to better estimate the temperature at the flux excess epoch. Moreover, it is the second 03fg-like SN Ia with oxygen features at late times, a highly unusual observation within the extended thermonuclear family. Apart from these observations, SN\,2021zny displays all the usual characteristics of the 03fg-like class, such as the high peak brightness compared to its (slow) decline rate, the UV-bright luminosity and blue early-time UV and optical colours (compared to normal SNe Ia), the low ejecta velocities and the persistent carbon features in its optical spectra. From all of the above, combined with the absence of hydrogen and helium features, the low ionization state and the low abundance of stable iron-peak elements, inferred from the late-time spectrum, we conclude that SN\,2021zny originated from a double-degenerate progenitor system, possibly a merger of two C/O WDs where the explosion occurs after the secondary WD is fully disrupted, from which some H/He-poor CSM is ejected at a close vicinity to the explosion site.

We note that it appears that every 03fg-like SN Ia that is discovered and monitored in the first $\sim1-3$ d from explosion shows this early flash, indicating a common explosion mechanism and progenitor system, with their peak-light and late-time diversities possibly explained by the variation of other model parameters. High-quality observations from very early to very late epochs, particularly in unexplored wavelength ranges, such as X-ray and near/mid IR would prove crucial in order to uncover the nature of these unique events. 

\section*{Acknowledgements}

We thank the anonymous referee for helpful comments that improved the clarity and presentation of this Paper.

GD and KM are supported by the H2020 European Research Council grant no. 758638.

AAM is partially supported by NASA grant 80NSSC22K0541.

MC acknowledges support from the National Science Foundation with grant numbers PHY-2010970 and OAC-2117997.

LG and TEMB acknowledge financial support from the Spanish Ministerio de Ciencia e Innovaci\'on (MCIN), the Agencia Estatal de Investigaci\'on (AEI) 10.13039/501100011033 under the PID2020-115253GA-I00 HOSTFLOWS project, from Centro Superior de Investigaciones Cient\'ificas (CSIC) under the PIE project 20215AT016, and by the program Unidad de Excelencia Mar\'ia de Maeztu CEX2020-001058-M. LG also acknowledges MCIN, AEI and the European Social Fund (ESF) "Investing in your future" under the 2019 Ram\'on y Cajal program RYC2019-027683-I.

MG is supported by the EU Horizon 2020 research and innovation programme under grant agreement No 101004719.

NI is partially supported by the Polish NCN DAINA grant No. 2017/27/L/ST9/03221.

MN is supported by the European Research Council (ERC) under the European Union's Horizon 2020 research and innovation programme (grant agreement No.~948381) and by a Fellowship from the Alan Turing Institute.

SS acknowledges support from the G.R.E.A.T. research environment, funded by {\em Vetenskapsr\aa det},  the Swedish Research Council, project number 2016-06012.

QW is supported in part by NASA grant 80NSSC22K0494, 80NSSC21K0242 and 80NSSC19K0112.

YY is supported by a Bengier-Winslow-Robertson Fellowship.

This work was funded by ANID, Millennium Science Initiative, ICN12\_009.

Based on observations obtained with the Samuel Oschin Telescope 48-inch and the 60-inch Telescope at the Palomar Observatory as part of the Zwicky Transient Facility project. ZTF is supported by the National Science Foundation under Grant No. AST-2034437 and a collaboration including Caltech, IPAC, the Weizmann Institute of Science, the Oskar Klein Center at Stockholm University, the University of Maryland, Deutsches Elektronen-Synchrotron and Humboldt University, the TANGO Consortium of Taiwan, the University of Wisconsin at Milwaukee, Trinity College Dublin, Lawrence Livermore National Laboratories, IN2P3, University of Warwick, Ruhr University Bochum and Northwestern University. Operations are conducted by COO, IPAC, and UW.

The ZTF forced-photometry service was funded under the Heising-Simons Foundation grant \#12540303 (PI: Graham).

The SED Machine is based upon work supported by the National Science Foundation under Grant No. 1106171.

The Liverpool Telescope is operated on the island of La Palma by Liverpool John Moores University in the Spanish Observatorio del Roque de los Muchachos of the Instituto de Astrofisica de Canarias with financial support from the UK Science and Technology Facilities Council.

Based on observations collected at the European Organisation for Astronomical Research in the Southern Hemisphere, Chile, as part of ePESSTO+ (the advanced Public ESO Spectroscopic Survey for Transient Objects Survey). ePESSTO+ observations were obtained under ESO program IDs 1103.D-0328, 106.216C, 108.220C (PI: Inserra). The Las Cumbres Observatory data have been obtained via OPTCON proposals (IDs: OPTICON 21B/001 and 22A/004). The OPTICON project has received funding from the European Union's Horizon 2020 research and innovation programme under grant agreement No 730890.

Based on observations made with the Nordic Optical Telescope, owned in collaboration by the University of Turku and Aarhus University, and operated jointly by Aarhus University, the University of Turku and the University of Oslo, representing Denmark, Finland and Norway, the University of Iceland and Stockholm University at the Observatorio del Roque de los Muchachos, La Palma, Spain, of the Instituto de Astrofisica de Canarias.

The data presented here were obtained in part with ALFOSC, which is provided by the Instituto de Astrofisica de Andalucia (IAA) under a joint agreement with the University of Copenhagen and NOT.

Some of the data presented herein were obtained at the W. M. Keck Observatory, which is operated as a scientific partnership among the California Institute of Technology, the University of California and the National Aeronautics and Space Administration. The Observatory was made possible by the generous financial support of the W. M. Keck Foundation.

The authors wish to recognise and acknowledge the very significant cultural role and reverence that the summit of Maunakea has always had within the indigenous Hawaiian community.  We are most fortunate to have the opportunity to conduct observations from this mountain.

We acknowledge the use of Weizmann Interactive Supernova Data Repository (WISeREP) maintained by the Weizmann Institute of Science computing center.

This work made use of the Heidelberg Supernova Model Archive (HESMA).

\section*{Data Availability}

The data analysed in this paper are available in the electronic edition and via the Weizmann Interactive Supernova Data Repository (WISeREP).



\bibliographystyle{mnras}
\bibliography{SN2021zny} 




\appendix

\section{Photometry and Spectroscopy}

\begin{table*}
\caption{Observed ground-based photometry of SN\,2021zny.}
\label{tab:phot_table}
\begin{tabular}{ccccccc}
    \hline
    MJD & Phase$^a$ & Telescope/Instrument & Filter & Brightness & Brightness error & Brightness upper limit$^b$ \\
     & (Rest-frame Days) &  &  & (AB mag) & (AB mag) & (AB mag) \\
    \hline
59472.38 & -25.40 & P48/ZTF & \textit{r} & & & 21.652 \\
59472.44 & -25.34 & P48/ZTF & \textit{g} & & & 21.735 \\
59474.33 & -23.50 & P48/ZTF & \textit{r} & & & 21.415 \\
59474.40 & -23.43 & P48/ZTF & \textit{g} & & & 21.827 \\
59476.37 & -21.51 & P48/ZTF & \textit{i} & & & 20.482 \\
59476.43 & -21.45 & P48/ZTF & \textit{r} & & & 20.940 \\
59476.50 & -21.38 & P48/ZTF & \textit{g} & 20.353 & 0.188 & \\
59477.29 & -20.62 & P48/ZTF & \textit{r} & 20.318 & 0.238 & \\
59477.35 & -20.55 & P48/ZTF & \textit{g} & 20.066 & 0.190 & \\
59478.28 & -19.65 & P48/ZTF & \textit{r} & & & 20.325 \\
59478.38 & -19.56 & P48/ZTF & \textit{g} & 20.146 & 0.336 & \\
59479.32 & -18.63 & P48/ZTF & \textit{i} & 19.415 & 0.244 & \\
59479.37 & -18.59 & P48/ZTF & \textit{r} & 19.299 & 0.116 & \\
59483.28 & -14.78 & P48/ZTF & \textit{g} & 17.207 & 0.040 & \\
59483.32 & -14.74 & P48/ZTF & \textit{i} & 17.364 & 0.052 & \\
59484.34 & -13.75 & P48/ZTF & \textit{r} & 16.941 & 0.032 & \\
59484.39 & -13.70 & P48/ZTF & \textit{g} & 16.903 & 0.032 & \\
59485.44 & -12.68 & ATLAS & \textit{o} & 16.880 & 0.020 & \\
59486.30 & -11.84 & P48/ZTF & \textit{r} & 16.682 & 0.031 & \\
59487.31 & -10.86 & P48/ZTF & \textit{i} & 16.736 & 0.036 & \\
59487.34 & -10.83 & P48/ZTF & \textit{r} & 16.449 & 0.029 & \\
59487.41 & -10.76 & P48/ZTF & \textit{g} & 16.352 & 0.029 & \\
59487.52 & -10.65 & ATLAS & \textit{o} & 16.570 & 0.010 & \\
59488.35 & -9.85 & P48/ZTF & \textit{r} & 16.327 & 0.029 & \\
59489.58 & -8.65 & ATLAS & \textit{o} & 16.370 & 0.010 & \\
59490.05 & -8.19 & LT/IO:O & \textit{u} & 16.230 & 0.060 & \\
59490.06 & -8.18 & LT/IO:O & \textit{z} & 16.770 & 0.040 & \\
59490.06 & -8.18 & LT/IO:O & \textit{i} & 16.580 & 0.040 & \\
59490.06 & -8.18 & LT/IO:O & \textit{g} & 16.060 & 0.050 & \\
59490.06 & -8.18 & LT/IO:O & \textit{r} & 16.190 & 0.030 & \\
59491.00 & -7.26 & NTT/SOFI & \textit{H} & 17.734 & 0.090 & \\
59491.00 & -7.26 & NTT/SOFI & \textit{J} & 17.009 & 0.050 & \\
59491.07 & -7.20 & LT/IO:O & \textit{g} & 15.970 & 0.050 & \\
59491.07 & -7.20 & LT/IO:O & \textit{u} & 16.160 & 0.030 & \\
59491.08 & -7.19 & LT/IO:O & \textit{i} & 16.510 & 0.040 & \\
59491.08 & -7.19 & LT/IO:O & \textit{r} & 16.100 & 0.040 & \\
59491.08 & -7.19 & LT/IO:O & \textit{z} & 16.680 & 0.050 & \\
59491.24 & -7.03 & P48/ZTF & \textit{r} & 16.041 & 0.029 & \\
59491.34 & -6.93 & P48/ZTF & \textit{i} & 16.442 & 0.033 & \\
59491.52 & -6.76 & ATLAS & \textit{o} & 16.210 & 0.010 & \\
59493.42 & -4.91 & ATLAS & \textit{o} & 16.150 & 0.010 & \\
59494.93 & -3.44 & LCO/Sinistro & \textit{u} & 16.162 & 0.024 & \\
59494.93 & -3.43 & LCO/Sinistro & \textit{g} & 15.783 & 0.011 & \\
59494.94 & -3.43 & LCO/Sinistro & \textit{r} & 15.966 & 0.010 & \\
59494.94 & -3.43 & LCO/Sinistro & \textit{i} & 16.377 & 0.012 & \\
    \hline
\multicolumn{7}{l}{$^a$ Relative to $B$-band maximum (MJD 59498.46)}\\
\multicolumn{7}{l}{$^b$ 3$\sigma$ upper limit}\\
\multicolumn{7}{l}{This table is available in its entirety in machine-readable form.}\\
  \end{tabular}
\end{table*}

\begin{table*}
\caption{Observed {\it Swift} photometry of SN\,2021zny.}
\label{tab:phot_table_swift}
\begin{tabular}{ccccc}
\hline
    MJD & Phase$^a$ & Filter & Brightness & Brightness error \\
     & (Rest-frame Days) & & (AB mag) & (AB mag) \\
    \hline
59487.83 & -10.35 & \textit{V} & 16.285 & 0.083  \\
59487.84 & -10.34 & \textit{UVM2} & 18.713 & 0.054  \\
59487.86 & -10.33 & \textit{UVW1} & 17.845 & 0.046  \\
59487.86 & -10.32 & \textit{U} & 16.498 & 0.029  \\
59487.86 & -10.32 & \textit{B} & 16.291 & 0.034  \\
59487.86 & -10.32 & \textit{UVW2} & 19.127 & 0.063  \\
    \hline
\multicolumn{5}{l}{$^a$ Relative to $B$-band maximum (MJD 59498.46)}\\
\multicolumn{5}{l}{This table is available in its entirety in machine-readable form.}\\
\end{tabular}
\end{table*}

\begin{table*}
\caption{Observed {\it TESS} 6h averaged photometry of SN\,2021zny.}
\label{tab:phot_table_tess}
\begin{tabular}{ccccc}
\hline
    MJD & Phase$^a$ & Brightness & Brightness error & Brightness upper limit$^b$ \\
     & (Rest-frame Days) & (AB mag) & (AB mag) & (AB mag) \\
    \hline
59473.71 & -24.10 &  &  & 19.831 \\
59473.87 & -23.94 &  &  & 20.252 \\
59474.12 & -23.70 &  &  & 20.114 \\
59474.37 & -23.45 &  &  & 20.187 \\
59474.62 & -23.21 &  &  & 20.218 \\
59474.88 & -22.96 &  &  & 20.200 \\
59475.12 & -22.72 &  &  & 20.132 \\
59475.37 & -22.48 &  &  & 20.449 \\
59475.62 & -22.24 &  &  & 20.399 \\
59475.87 & -21.99 &  &  & 20.438 \\
59476.12 & -21.75 &  &  & 20.188 \\
59476.37 & -21.51 &  &  & 20.202 \\
59476.62 & -21.26 &  &  & 20.315 \\
59476.87 & -21.02 & 20.089 & 0.350 &  \\
59477.12 & -20.78 & 19.638 & 0.202 &  \\
59477.37 & -20.53 &  &  & 20.276 \\
59477.61 & -20.30 & 20.164 & 0.361 &  \\
59477.87 & -20.05 &  &  & 20.699 \\
59478.12 & -19.80 & 20.341 & 0.362 &  \\
59478.37 & -19.56 & 20.223 & 0.318 &  \\
59478.62 & -19.32 &  &  & 20.249 \\
59478.87 & -19.07 & 19.865 & 0.250 &  \\
59479.12 & -18.83 & 19.737 & 0.198 &  \\
59479.37 & -18.59 & 19.283 & 0.182 &  \\
59479.62 & -18.34 & 19.003 & 0.107 &  \\
59479.87 & -18.10 & 18.861 & 0.093 &  \\
    \hline
\multicolumn{5}{l}{$^a$ Relative to $B$-band maximum (MJD 59498.46)}\\
\multicolumn{5}{l}{This table is available in its entirety in machine-readable form.}\\
\end{tabular}
\end{table*}

\begin{table*}
\caption{Observing log of the optical spectra of SN\,2021zny.}
\label{tab:spec_log}
\begin{tabular}{lcccc}
\hline
    Obs Date & Phase$^a$ & Telescope/Instrument & Slit Width & Grism/Grating \\
    (UT) & (Rest-frame Days) & & \\
\hline
2021-09-27 & -13.90 & P60/SEDM & -- & -- \\
2021-09-27 & -13.82 & P200/DBSP & & \\
2021-09-27 & -13.69 & P60/SEDM & -- & -- \\
2021-09-30 & -10.84 & P60/SEDM & -- & -- \\
2021-10-03 & -8.03 & NTT/EFOSC & 1\farcs0 & Gr\#11+Gr\#16 \\
2021-10-03 & -8.00 & LT/SPRAT & 1\farcs8 & 600 lines/mm, Red \\
2021-10-04 & -7.08 & P60/SEDM & -- & -- \\
2021-10-04 & -6.65 & Keck/LRIS & 1\farcs0 & 400/3400+400/8500 \\
2021-10-06 & -5.11 & NTT/EFOSC & 1\farcs0 & Gr\#11+Gr\#16 \\
2021-10-09 & -1.45 & LT/SPRAT & 1\farcs8 & 600 lines/mm, Red \\
2021-10-14 & 2.68 & NTT/EFOSC & 1\farcs0 & Gr\#11+Gr\#16 \\
2021-10-25 & 14.15 & LT/SPRAT & 1\farcs8 & 600 lines/mm, Red \\
2021-10-26 & 14.32 & NTT/EFOSC & 1\farcs0 & Gr\#11+Gr\#16 \\
2021-10-30 & 18.29 & P60/SEDM & -- & -- \\
2021-10-30 & 19.25 & Magellan/LDSS3 & 1\farcs0 & VPH-ALL \\
2021-11-04 & 23.09 & NTT/EFOSC & 1\farcs0 & Gr\#11+Gr\#16 \\
2021-11-26 & 44.48 & NTT/EFOSC & 1\farcs0 & Gr\#11+Gr\#16 \\
2021-11-30 & 48.32 & P60/SEDM & -- & -- \\
2021-12-01 & 49.21 & P200/DBSP & & \\
2021-12-08 & 56.85 & LT/SPRAT & 1\farcs8 & 600 lines/mm, Red \\
2021-12-22 & 70.62 & NOT/ALFOSC & 1\farcs3 & grism \#4 \\
2022-01-03 & 81.57 & P60/SEDM & -- & -- \\
2022-01-04 & 82.43 & NTT/EFOSC & 1\farcs0 & Gr\#11+Gr\#16 \\
2022-01-07 & 85.41 & P60/SEDM & -- & -- \\
2022-01-12 & 90.30 & P60/SEDM & -- & -- \\
2022-01-21 & 99.11 & P60/SEDM & -- & -- \\
2022-01-26 & 103.88 & P60/SEDM & -- & -- \\
2022-02-15 & 124.02 & NOT/ALFOSC & 1\farcs0 & grism \#4 \\
2022-08-29 & 312.59 & Keck/DEIMOS & -- & LVMslitC \\
\hline
\multicolumn{5}{l}{$^a$ Relative to $B$-band maximum (MJD 59498.46)}\\
\end{tabular}
\end{table*}


\bsp	
\label{lastpage}
\end{document}